\documentclass[table]{ccjnl}
\usepackage{graphicx}
\usepackage{subfloat}
\usepackage{epstopdf}
\usepackage{lipsum,amsmath}
\usepackage{amsmath}
\usepackage{cuted}
\usepackage{bm}
\usepackage{enumitem}
\usepackage{color}

\title{Computing Power Network: A Survey}
\author{Yukun Sun \inst{1}, Bo Lei \inst{2,*}, Junlin Liu \inst{1}, Haonan Huang \inst{1}, Xing  Zhang\inst{1,*},  Jing Peng \inst{3}, Wenbo Wang  \inst{1} 
}
\receiveddate{Sep. 25, 2019}
\reviseddate{Mar. 24, 2020}
\Editor{name}

\address[1]{School of Information and Communications Engineering, Beijing University of Posts and Telecommunications, Beijing 100876, China}
\address[2]{Research Institute of China Telecom Co., Ltd., Beijing 102209,China}

\address[3]{Beijing Branch of China Telecom Co., Ltd., Beijing 100032,China}
\address[*]{The corresponding authors}

\begin{document}
\maketitle

\begin{abstract}
 With the rapid development of cloud computing, edge computing, and smart devices, computing power resources indicate a trend of ubiquitous deployment. The traditional network architecture cannot efficiently leverage these distributed computing power resources due to computing power island effect. To overcome these problems and improve network efficiency, a new network computing  paradigm is proposed, \textit{i.e.}, Computing Power Network (CPN). Computing power network can connect ubiquitous and heterogenous computing power resources through networking to realize computing power scheduling flexibly. In this survey, we make an exhaustive review on the state-of-the-art research efforts on computing power network. We first give an overview of computing power network, including definition, architecture, and advantages. Next, a comprehensive elaboration of issues on computing power modeling, information awareness and announcement, resource allocation, network forwarding, computing power transaction platform and resource orchestration platform is presented. The computing power network testbed is built and evaluated. The applications and use cases in computing power network are discussed. Then, the key enabling technologies for computing power network are introduced. Finally, open challenges and future research directions are presented as well.
\end{abstract}

\section{INTRODUCTION}\label{s1}
With the advancement of the fifth generation (5G) mobile  communication network and  the rapid development of artificial intelligence (AI) technology, there are a wide variety of new applications emerging, such as Augmented Reality (AR), Virtual Reality (VR), Industrial Internet, Internet of Vehicles (IoV), Internet of Things (IoT), etc. According to a recent report from Cisco \cite{Cisco1}, the number of devices connected to IP network  will be more than three times the global population by 2023. There will be 29.3 billion networked devices by 2023, up from 18.4 billion in 2018. These emerging various applications and the explosive growth of networked devices  bring huge challenges to the network transmission ability and computing processing ability. The concept of mobile edge computing (MEC) was proposed and renamed as multi-access edge computing (MEC). MEC which can  sink  cloud computing resources into network edge such as base station or gateway, is a promising approach for limited computing resources in end devices and busy communication link. 

With the maturity of MEC technology,  the combination of MEC and AI makes edge intelligence emerge our view. Edge intelligence refers to the deployment of AI algorithms, technologies, and products in the edge of wireless network. It is an open platform which integrates core capabilities such as network, computing, storage and applications. The proposal of edge intelligence emphasizes being close to the source of data and deploying intelligence in edge devices to reduce the delay in the delivery of intelligent cloud computing services and provide users with intelligent services faster and better. Its vision is to realize intelligent perception of wireless conditions, intelligently rapid decision-making and real-time response. There are growing computing  intensive and delay insensitive services in the scenario of edge intelligence. Large-scale applications of AI algorithm across various industries need enormous computing power resources. However, the computing power of a single computing node is limited due to the computing power  bottleneck of hardware infrastructures. Thus, it is necessary to deploy ubiquitous computing power resources. 

With the large and ubiquitous deployment of computing power resources pools and intelligent end devices, it becomes more convenient  and faster to enable users to access and leverage the massive distributed computing resources \cite{WSPN}. However, the existing computing power resources pools and end devices with computing capability are scattered in location. There is no  efficient coordination approach among edge computing nodes or between  edge computing nodes and cloud computing nodes, which leads to the computing resources utilization being pretty low. On the other hand, the processing latency is very high for some computing intensive tasks allocated in a single computing node with limited computing power resources,  i.e., computing power island effect, which may not satisfy the users' requirements. What's worse, there will exist  load unbalancing due to space-time uneven distribution of computation  task. 

The development of network has laid the foundation  for the flexible provision of multi-party computing resources.  With the advancement of emerging network technologies such as software defined network  (SDN), network function virtualization (NFV), and network slicing, etc., computing power resources  can be connected  efficiently  and dynamically through network. 

To address the problems above, the concept of computing power network (CPN) is proposed to coordinate ubiquitously distributed computing resources. The core purpose of CPN is to schedule computing power intelligently, on demand, dynamically and efficiently, breaking  the island effect of computing power in the era of MEC. A comparison of different computing paradigms  from cloud computing to computing power network is summarized in Table \ref{table0}. Firstly, there are many distributed computing nodes deployed close to users in CPN, which makes  computing tasks no longer need to be transmitted over long distances compared with cloud computing. Moreover, ubiquitous computing resources can be connected and coordinated through network in CPN, which will provide  more abundant computing resources for computing tasks compared with isolated MEC node. Therefore, CPN will bring lower transmission latency compared with cloud computing and lower computing latency compared with MEC, which will enable ultra-low latency task processing. In addition, MEC did not effectively reduce the latency during actual testing in fact  because of solidified network architecture. Secondly, distributed computing nodes can aware more information about computing tasks and network condition. Meanwhile, the information can be shared among  ubiquitous computing nodes in CPN, which will make CPN have robust awareness capability. Thirdly, in terms of application scenarios, some  major companies provide computing services for users in the era of cloud computing, such as Ali Cloud and Azure. In the era of MEC, operators join in deploying computing nodes at the edge of the network, which also provides computing services for users that are mainly IoT devices with limited energy and computing resources. However, in the era of CPN, every networked device with computing capability can become a computing power provider. For example, each vehicle in IoV will be not only a consumer but also a provider. Finally, both cloud computing and MEC are isolated carriers of computing resources and storage resources. CPN will upgrade current network architecture to leverage these isolated resources more effectively.

\begin{table*}[htbp]
\centering
\caption{Comparison of different computing paradigms.}
\begin{tabular}{|p{5cm }|  p{3.5cm}| p{3.5cm}|p{3.5cm}| }\hline
\textbf{Item} & \textbf{Cloud Computing  } & \textbf{MEC} & \textbf{CPN} \\\hline
\textbf{Proposed time/ year  } & 2006 & 2014 & 2019 \\\hline
\textbf{Core idea} & Provide computing service in cloud center. & Sink the computing resources into network edge.  & Connect and coordinate distributed computing resources through network.   \\\hline
\textbf{Cooperation among nodes } & No & No &  Yes   \\ \hline 
\textbf{Latency} & High & Low &Ultra-low  \\ \hline 
\textbf{Computing capability } & High & Limited & High  \\ \hline 
\textbf{Awareness ability} & No & Limited  & High   \\ \hline
\textbf{Key technologies}  & Virtualization   & Computing offloading  & Computing power routing   \\\hline
\textbf{Application scenarios  }  & Ali Cloud, Azure  & IoT & IoV  \\\hline
\textbf{Weaknesses  }  & Network bottleneck & Computing power island effect & Huge network  equipment upgrade cost \\\hline
\end{tabular}
\label{table0}
\end{table*}

\subsection{related work }
Computing power network has been drawing much attention since proposed in 2019. In this section, the related work of computing power network is surveyed from  standardization  and academic research  aspects as follows.

In terms of standardization, the Internet Engineering Task Force (IETF) established Computing in Network Research Group (COINRG) in February, 2019, which is devoted to studying deep integration of computing and networking. Computing in network means that network infrastructures will have both forwarding and computing abilities. IETF published Computing First Network (CFN) scenarios and requirements \cite{IETF1}, framework of CFN \cite{IETF2}, and the report on CFN \cite{IETF3}. Meanwhile, network operators and equipment vendors like Huawei and so on enterprises propel Y.CPN-arch \cite{ITU2}, Y.IMT2020-CAN-req \cite{ITU1} standardization in  International Telecommunication Union Telecommunication Standardization Sector (ITU-T) Study Group 13 (SG13). In July, 2021, Y.2501 formly Y.CPN-arch is proposed \cite {ITU3}. ITU-T Study Group 11 (SG11) started Q.CPN standardization about signaling  requirements for computing power network \cite{ITU4} and Q.BNG-INC about requirements and signaling of intelligence control for the border network gateway in computing power network \cite{ITU5}. Broadband Forum (BBF) started Metro Computing Network (MCN) project (SD-466) devoted to studying computing power network in Metro network in 2109, and develops to  MCN draft (WT-466) project \cite{BBF1} looking  at the cloud computing locations, particularly edge computing locations in 2021.  China Communications  Standards  Association (CCSA) and Network 5.0 Industry and Technology Innovation Alliance also lead computing power network standardization research work and established computing power network special work group at  Network 5.0 Technology Standardization Promotion Committee (CCSA TC614) \cite{CCSA}. Moreover, CCSA TC3 released the report of requirements and architecture for computing power network and promoted related research work from 2019 to present. There are many white papers released  to explore computing power network also named computing-aware network (CAN) since the concept of computing power network is proposed \cite{w3,w4,w1,w2}. The concept of terminals computing aware network (TCAN) is also proposed \cite{tcpn} as well as  the authors propose four special characters and three network forms.

In terms of academic research, in \cite{D1}, the concept of computing power network is first proposed. The author studied the computing power network scheme for cloud, network, edge collaboration and analyzed its characteristics in computing power abstraction, service guarantee, unified control, and elastic scheduling. The work of \cite {L1}  illustrated the technical solution of computing power network can efficiently meet the multi-level deployment and flexible scheduling needs of the future 6G business for computing, storage and network. The basic architecture and the process of working in CFN is presented  in \cite{1121}.  A computing-networking scheme based on the deep fusion of cloud, edge and network, called compute first networking  is introduced in \cite{1122}. It elaborated CFN technology framework and CFN routing protocol. In \cite{1123}, the author elaborated the main technology challenges and prospects for edge computing and computing power network. The distributed architecture of edge computing makes it easier to be attacked. In addition, the more intelligent the client, the easier the system is  attacked by malicious software. In the future, we need to promote the research work of computing power network from three aspects, i.e., new network architecture from computing and networking management separately to computing and networking management uniformly, new network protocol from choosing  the best transmission path to jointly choosing  the best transmission path and computing nodes, new metric standard from network metric to jointly metric of network and computing power \cite{1124}. In \cite{1125}, the author proposed a novel network virtualization architecture based on the convergence of computing, storage, and transport multi-dimensional resources in the trend of general basic resources, intelligent network services, and diversified business demand in the future network. 

\begin{figure*}[htbp]
\centering 
\includegraphics[width=\linewidth,scale=1.00 ]{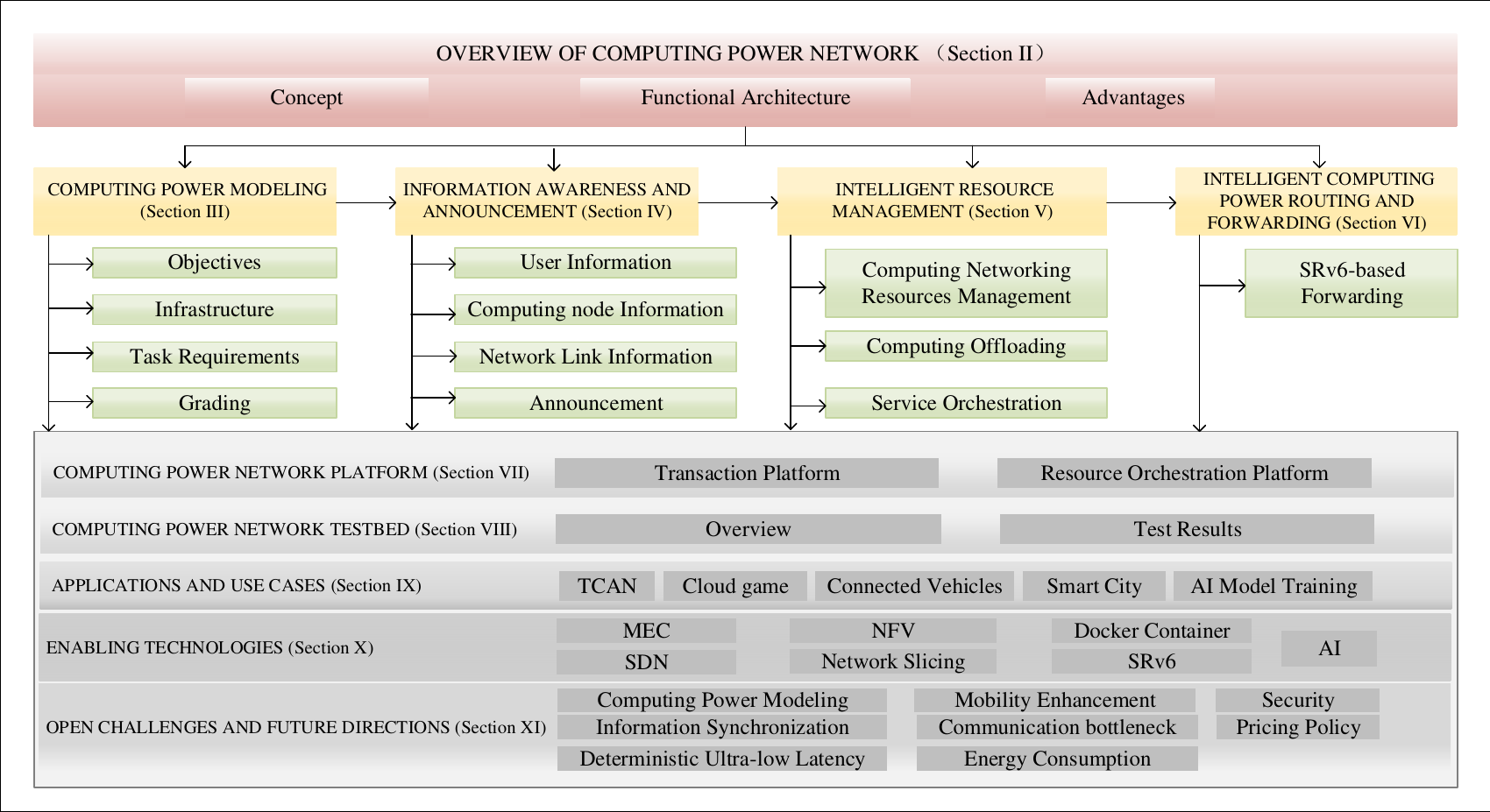}
\caption{Road map of computing power network survey.}
\label{F0}
\end{figure*}

\subsection{survey novelty and contributions}
To the best of our knowledge, different from  the previous works, this survey is the first to introduce computing power network comprehensively  from  definition, architecture, advantages, functional models, platform, enabling technologies, applications and  use cases, as well as open challenges and future directions. The main contributions of this article are summarized as follows.
\begin{itemize}
    \item Overview of computing power network functional architecture  is conducted, and the advantages of computing power network is stated.
    
    \item A comprehensive survey of key aspects of computing power network is presented. The ways of computing power modeling  are introduced. The information perception  and the ways of information announcement are summarized. The optimal strategies for intelligent resource management are surveyed and the network forwarding based on segment routing IPv6 (SRv6) is elaborated. Computing power network transaction platform and resource orchestration platform are proposed. 
    
    \item The computing power network testbed is built and evaluated. The applications and use cases of computing power network are comprehensively summarized. The key enabling technologies for computing power network are pointed, including multi-access edge computing, software defined network, network function virtualization, network slicing, docker container, segment routing IPv6, and artificial intelligence. 
    
    \item The open challenges and future research directions related to computing power network are identified, such as computing power modeling, information synchronization, mobility enhancement, energy consumption, deterministic ultra-low latency and so on.
\end{itemize}

\subsection{survey structure}
The remainder of this article is organized as follows.
\begin{itemize}
    \item Section \ref{s2}: The concept of computing power network is elaborated. We introduce the functional architecture of computing power network. The advantages of computing power network are given. 
    \item Section \ref{s3}: We survey the computing power modeling in computing power network including computing power measurement objectives, infrastructure computing power modeling, task requirements computing power modeling and proper computing power grading. 
    \item Section \ref{s4}: We survey available and valuable information to aware in computing power network, including user information, computing node information and network link information. Three ways of information announcement are also elaborated in this section. 
    \item Section \ref{s5}: Intelligent resource management solutions are surveyed in this section including computing-networking resources registration and management, computing offloading, and service orchestration. 
    \item Section \ref{s6}: Intelligent network forwarding in computing power network is introduced. 
    \item Section \ref{s7}: We introduce the computing power network transaction platform and computing power network  resource orchestration platform.
     \item Section \ref{sA1}: We overview the actual computing power network  testbed built and give some evaluation results.
    \item Section \ref{s8}: We analyze possible applications and use cases with computing power network. 
    \item Section \ref{s9}: We survey several key enabling technologies for computing power network. 
    \item Section \ref{s10}: Open challenges and future directions based on computing power are proposed. 
    \item Section \ref{s11}: The key ideas from this survey are concluded.
\end{itemize}
For convenience, a summary of all abbreviations is shown in  Table ~\ref{table1}. A road map of this survey is illustrated as shown in Figure \ref{F0}.

\begin{table}[htbp]\scriptsize
\centering
\caption{Summary  of abbreviations.}
\begin{tabular}{p{1.7cm } p{6.1cm}  }\hline
5G & The Fifth Generation Mobile Communication System\\
B5G/6G & Beyond 5G / The Sixth Generation \\
AI & Artiﬁcial Intelligence\\
AIoT & Artificial Intelligence \& Internet of Things\\
AR & Augmented Reality\\
BAIR & Berkeley AI Research\\
BBF &  Broadband Forum\\
BGP & Border Gateway Protocol\\
BPN & Back Propagation Network \\
CAN &  Computing-aware  Network \\
CCSA & China Communications Standards Association \\
CCF & Chinese Computer Society\\
CFN & Computing First  Network \\
CNN & Convolutional Neural network \\
COINRG & Computing in Network  Research  Group\\
CPU & Central Processing Unit\\
CPN & Computing Power Network\\
DDPG & Deep Deterministic Policy Gradient\\
DetNet & Deterministic Networking\\
DNN & Deep Neural Network\\
DPU & Deep Learning Processing Unit\\
DQN & Deep Q-learning Network\\
DRL & Deep Reinforcement Learning\\
DTSO & Distributed-Two-Stage Offloading\\
ETSI & European Telecommunication Standards Institute\\
FIB & Forwarding Information Base\\
FL & Federal Learning \\
FLOPS & Floating-Point Operations Per Second\\
FPS & Frames Per Second\\
GPU & Graphics Processing Unit\\
G-SRv6 & Generalized Segment Routing Over IPv6\\
IETF & Internet Engineering Task Force \\
IGP & Interior Gateway Protocol\\
IoT & Internet of Things\\
IoV & Internet of Vehicles\\
ITU-T & International  Telecommunication  Union Telecommunication Standardization Sector\\
K3S & Lightweight Kubernetes\\
K8S & Kubernetes\\
LSTM & Long  Short  Term  Memory  Network \\
MAPS & Mean Accuracy-Guaranteed Processing Speed\\
MCN & Metro Computing Network\\
MEC & Mobile/Multi-access Edge Computing \\
MIPS & Million Instructions Per Second\\
NFV & Network  Function Virtualization \\
NFV MANO & Network Function Virtualization Management and Orchestration\\
NFVO & Network Function Virtualization Orchestrator\\
NPU & Neural Network Processing Unit\\
PDV & Packet Delay Variation\\
RIB & Routing Information Base\\
RNN & Recurrent Neural Network\\
SDN &  Software Defined Network\\
SFC & Service Function Chaining\\
SLA & Service Level Agreement\\
SR & Segment Routing\\
SRv6 & Segment Routing IPv6\\
TE & Trafﬁc Engineering\\
TOPS & Trillion Operations Per Second\\
TPU & Tensor Processing Unit\\
VNF & Virtual Network Function\\
VNFPRA & Virtual Network Function Placement and Resource Allocation\\
VPN & Virtual Private Network\\
VR & Augmented Reality\\\hline
\end{tabular}
\label{table1}
\end{table}

\section{OVERVIEW OF  Computing Power Network}\label{s2}
With the deployment of edge computing servers  and intelligent terminal devices, computing power  resources  present a ubiquitous  and heterogeneous  deployment trend. The traditional mobile edge network  architecture cannot make full use of massive distributed computing power  resources  any more. The next generation network  will converge distributed computing  power resources and network deeply as well as coordinate these computing power  resources. In this section, we will firstly explain what is computing power network. Then the architecture of  computing power network is presented. At last, we discuss the advantages of  computing power network.
\subsection{what is computing power network}\label{s21}
The core idea of  computing power network is to connect the distributed computing nodes. Computing power network can dynamically and timely perceive user demand and  multi-dimensional resources such as application, network resources, computing power resources, and  storage resources.  Computing power network jointly allocate computing power resources and network resources, and coordinately schedule computing tasks so that the application can call computing power resources in different places on-demand and timely. Computing power network is a network where computing power resources can be perceived, allocated and scheduled intelligently, which can provide optimal user experience and computing and network resource utilization ratio to meet the computing power requirements of new applications in beyond 5G or 6G. Computing power network is a new paradigm of deep integration of end-edge-cloud-network, and is also a  new stage in the evolution of edge computing to ubiquitous computing \cite{223}.

Computing power network will transmit not only data in bit but also ubiquitous and distributed computing resources compared with the existing networks. Computing power network is essentially a new network architecture jointly driven by AI, edge computing and new generation mobile communication network. Meanwhile, computing power network is still also based on edge computing and the existing network. Computing power network will connects ubiquitous computing resources in edge computing through updated network architecture with new enabling technologies.

\begin{figure*}[htbp]
\centering 
\includegraphics[width=\linewidth,scale=1.00 ]{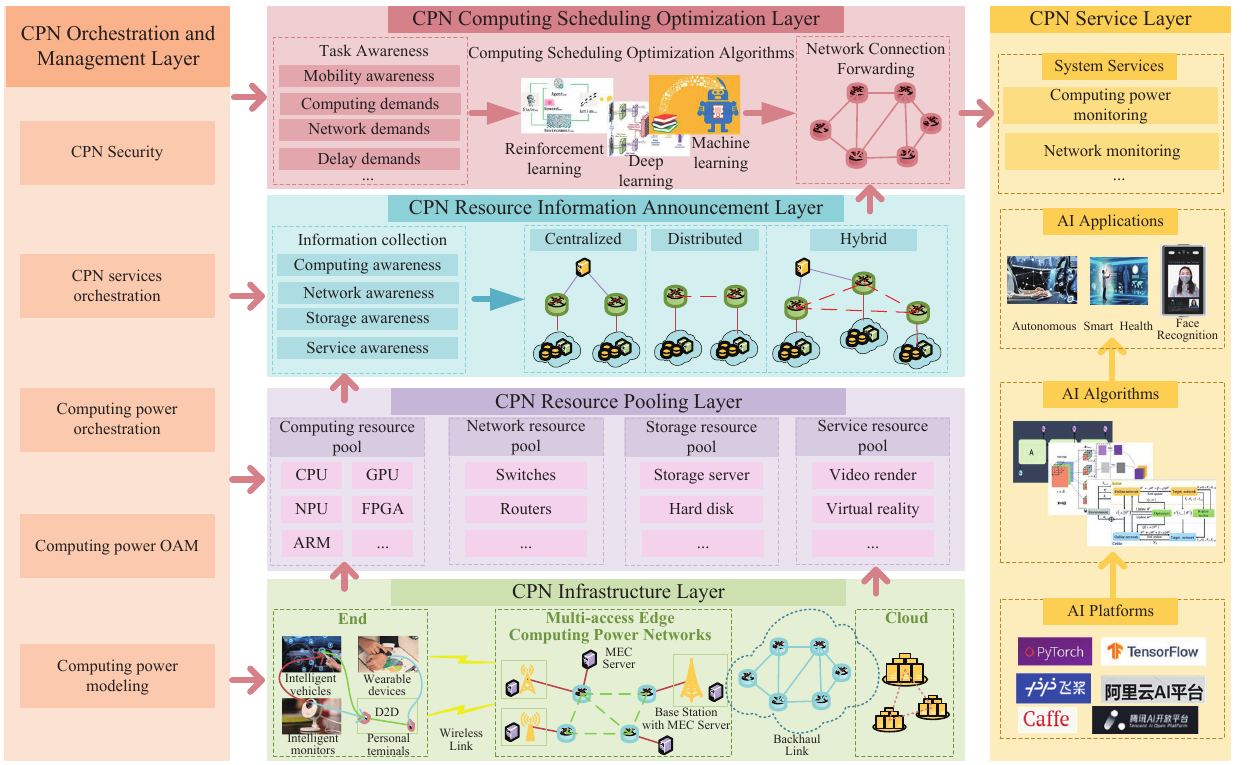}
\caption{Functional architecture of computing power network.}
\label{F21}
\end{figure*}

\subsection{functional architecture of computing power network}\label{s22}
In computing power network, it is most important to provide integrated computing services for users and achieve intelligent networking management. In fact, resources and services in computing power network are dependent on underlying hardware infrastructures enabling the whole system. How to model, abstract and allocate these computing resources, as well as achieve flexible and intelligent networking management has an important influence on the performance of computing power network. To achieve ubiquitous computing power and services perception, connectivity and collaborative scheduling, in this paper, we show the computing power network functional framework, as shown in Figure ~\ref{F21}, composed  of CPN infrastructure layer, CPN resource pooling layer, CPN resource information announcement layer, CPN computing scheduling optimization layer, CPN service layer as well as  CPN orchestration and management layer to be elaborated  as follows. 

\noindent\textbf{CPN Infrastructure Layer:} The CPN infrastructure layer is composed of computing infrastructures and network infrastructures. With the development of B5G and edge computing, computing power infrastructures are migrating from clouds to network edges and end devices with low latency, efficient mobile broadband, and high privacy. Computing power are deployed ubiquitously in end-edge-cloud. End devices, such as intelligent vehicles, wearable devices, intelligent monitors, mobile phones and laptops personal terminals, multi-access computing power network  edge, such as base stations, MEC servers and edge gateways, as well as clouds jointly provide computing power in computing power network. 

Network infrastructures in computing power network include wireless or wired access network, such as  WiFi, 5G base stations, 4G base stations, intelligent gateways, smart routers, optical fiber, and so on. The  ubiquitous connections among heterogenous computing devices will be supported by these network infrastructures to achieve computing power scheduling flexibly, intelligently, and on demand. 

\noindent\textbf{CPN Resource Pooling  Layer:} In this layer, multi-level computing resources and storage resources, ubiquitous network resources are modeled with general standard, abstracted and pooled, as well as the diverse services are deployed dynamically. CPN resource pooling layer need to perceive physical computing, storage, and network resources from CPN infrastructure layer,  while modeling and pooling the heterogenous and ubiquitous resources into computing resource pool, network resource pool and service resource pool. Moreover, CPN  resource pooling layer need to pool microservices provided to achieve flexible and dynamic deployment. Because of the heterogeneity of the underlying hardware infrastructures and the computing power from scattered computing providers, computing power modeling will be a major concern. At the same time, the privacy and transaction rules  of the resource pools are also considered to be very significant for computing power network. 

\noindent\textbf{CPN Resource Information Announcement  Layer:} This layer includes information collection from the CPN resource pooling layer and information synchronization in the whole computing power network. The interface in this layer can obtain the available computing, storage, network and service resources information by awaring  the resources of CPN resource pooling layer, which is instructive for diversified routing and forwarding strategies. Meanwhile, information about computing and network resources needs to be synchronized by the three main ways, i.e., centralized information synchronization, distributed information synchronization or hybrid information synchronization.

\noindent\textbf{CPN Computing Scheduling Optimization  Layer:} Different applications have different resources requirements. The requirements are grouped into multiple classes according to their mobility awareness, computing demands, network demands, delay demands and so on. According to these demands from users, combined with the resources information from CPN resource information announcement layer, CPN computing scheduling optimization layer will make intelligent scheduling decisions and resources allocation strategies with computing scheduling optimization algorithms, such as reinforcement  learning, deep learning, machine  learning, and so on. The optimized objectives are to achieve the shortest latency of task processing, the most efficient computing and network resources utilization, the flexibility and intelligence for network management, and the most revenue for computing power providers.

With the optimal task scheduling decisions and resources allocation strategies, CPN computing scheduling optimization layer also need to  achieve network connection forwarding, which can offload computation tasks to corresponding computing nodes to execute. The network connection requires not only the traditional communication pipeline, it may but also require the deployment of corresponding network elements, such as 5G UPF, vBRAS, vCPE and other access control network elements, according to network connection requirements. How to achieve fast and reliable tasks forwarding is a major concern in  CPN computing scheduling optimization layer, which will be introduced in detail in the later section. 

\noindent\textbf{CPN Service  Layer:} Because the future applications are mainly the computing intensive and delay insensitive AI applications, CPN service layer need to  realize the optional and various AI algorithms and provide AI platforms required by user tasks \cite{y71}. According to users' tasks  requirements, computing power network system provides  the proper AI algorithms,  such as face recognition using convolution neural network (CNN), mobility prediction using long short term  memory network (LSTM), text recognition using back propagation network (BPN) and so on. Moreover, there are many emerging AI platforms to be deployed in this layer, such as Pytorch developed by  Facebook, Tensorflow developed by Google, PaddlePaddle developed by Baidu, Caffe  developed by Berkeley AI Research (BAIR) and by community contributors, Ali cloud AI platform, Tencent AI open platform, and so on. The emerging AI  services will be executed in this layer, with the allocated computing and network resources from CPN resource pooling layer, the various and optional AI algorithms and the provided integrated AI platforms. The services could be divided into two classes. On 
the one hand, system services mainly for running computing power network system, such as computing power monitoring, network monitoring, and so on. On the other hand, AI applications mainly for users' service requirements, such as autonomous, smart health, face recognition, and so on. In summary, CPN service layer will deploy various AI services to satisfy emerging new applications requirements. 

\noindent\textbf{CPN Orchestration and Management  Layer:} This layer can realize computing power modeling, computing power OAM, computing power orchestration,  service orchestration, security for computing power network.
The computing power modeling module is used to measure heterogenous and ubiquitous resources and users' resource requirements  with general standard. Computing power OAM realizes the operation, administration and maintenance. Computing power orchestration module is in charge of the orchestration and management of CPN resources. CPN services orchestration module is in charge of the orchestration and management of CPN microservices. The CPN security module  is responsible for computing power transaction security and other threats. 

\begin{figure}[htbp]
\centering 
\includegraphics[width=\linewidth,scale=1.00 ]{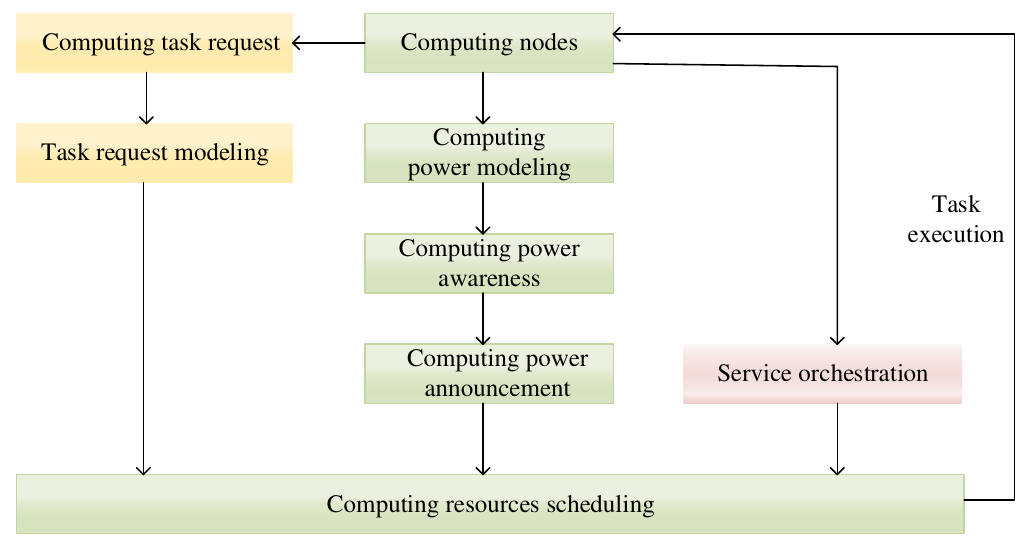}
\caption{The procedure of operating a computing task.}
\label{Ftask}
\end{figure}

As shown in Figure ~\ref{Ftask}, the procedure of operating a computing task for a user in computing power network is presented based on key function layers above. In terms of the computing task, the computing task first initiates an access request to CPN. Then, task request modeling microservice deployed in CPN will measure the computing power request of the computing task. In term of computing nodes, CPN first measure the computing power of every computing nodes. Then, CPN resource information announcement layer will aware computing power information and interact with other computing nodes for information interaction. Meanwhile, CPN orchestration and management layer need to deploy service for every computing nodes. The task request, computing nodes information and service deployment information will be sent to CPN computing scheduling optimize layer to make computing offloading decisions. Then, the computing task will be routed to the specific computing nodes to execute. Finally, the execution result will return to the user.

\subsection{advantages of computing power network}\label{s23}
Compared to current mobile edge network architecture, researchers found  computing power network have advantages in various aspects. Next, we will discuss them in detail. 
\begin{itemize}
\item[1)] Ultra-low latency 

The computing power requirements of user tasks, network condition and computing nodes resource state can  be perceived in real time based on corresponding protocol in  computing power network. Task scheduling can be executed intelligently and fast for computing intensive and delay  insensitive applications based on real-time information.  Moreover, ubiquitous computing resources in end-edge-cloud  can  be  coordinated  efficiently to provide ultra-low latency for user tasks.

\item[2)] High ubiquitous resources utilization 

Computing power network can collaborate  ubiquitous and heterogeneous  computing resources deployed  in end-edge-cloud based on ubiquitous network  connections. Therefore, computing intensive and delay insensitive tasks can be executed efficiently with flexible computing power scheduling. For example \cite{231}, there exist three computing nodes, one of the computing nodes with three computation tasks, however, the others without any computation task. Therefore, load unbalance causes resources under-utilized. However, the three tasks can be scheduled to other computing nodes to execute intelligently in computing power network with high ubiquitous resources utilization.

\item[3)] Consistency of user experience 

Computing power network can aware ubiquitous computing and services so that it is unnecessary for users to know the location and status of computing resources deployed. Computing and network resources in computing power network are scheduled coordinately to ensure the consistency of user experience. 

\item[4)] Flexible and dynamic service scheduling 

Services can be deployed dynamically at proper computing nodes with virtual network technology in computing power network. According to real-time network condition and computing resources status, particularly service level agreement (SLA), computing  power network can provide dynamic service deployment to ensure the user experience. 

\end{itemize}

\section{COMPUTING POWER MODELING IN Computing Power Network}\label{s3}
With the deployment of a lot of intelligent end devices and edge servers, computing power presents a general and heterogeneous deployment trend. How to measure and model computing power is the basis of computing power scheduling, computing power management and computing power transaction in computing power network.
\subsection{computing power measurement objectives  }\label{s31}
To provide differentiated services according to customers computing power requirements, it is necessary to quantify computing power. According to general computing power measurement objectives, it is possible to evaluate the existing computing power of the computing power resource pool and the customers computing power requirements. Computing power usually can be divided into logical computing power, parallel computing power and neural network computing power based on different algorithms and the type of computation data. 

\noindent\textbf{Logical computing power:} Logical computing power is a kind of general computing power. Central Processing Unit (CPU) is the representative of logical computing power. 70\% of the transistors are used to build the Cache, and there are some control units with few calculation units for logic control operations. Therefore, CPU computing power is limited for large-scale parallel computation, but it is most useful for logic control operations. TOPS (the processor performs a trillion operations per second) is usually  used to measure logical computing power.

\noindent\textbf{Parallel computing power:} Parallel computing  is a type of computing architecture in which several processors execute or process an application or computation simultaneously. Parallel computing power is  an efficient computing power specially to handle the applications in which the computation data type is unified, such as text processing, voice processing and video processing. Graphics Processing Unit (GPU) is the typical hardware chip representative. GPU is suitable for large-scale parallel computing, mainly used in big data, background server and  image processing. FLOPS (floating point computing power) is usually used to measure parallel computing power. 

\noindent\textbf{Neural network computing power:} Neural network  computing power is mainly used for computing intensive applications, such as  machine learning, deep learning and deep reinforcement learning, etc. Deep learning Processing Unit (DPU), Neural network Processing Unit (NPU), and Tensor Processing Unit (TPU) are the typical hardware chips representatives for emerging AI applications acceleration. Currently, FLOPS is also usually used to measure neural network computing power.

\begin{table*}[htbp]
\centering
\caption{Comparison of different computing power.}
\begin{tabular}{|p{5cm }|  p{3.5cm}| p{3.5cm}|p{3.5cm}| }\hline
\textbf{Item} & \textbf{Logical Computing Power } & \textbf{Parallel Computing Power} & \textbf{Neural Network Computing Power} \\\hline
\textbf{Typical chips  } & CPU & GPU & NPU,TPU \\\hline
\textbf{The unit of measurement} & TOPS & FLOPS  & FLOPS \\\hline
\textbf{Characteristics } & Logical control & Big data parallel processing with unified data type &  Processing AI tasks   \\ \hline 
\textbf{Application scenarios} & Weapons and equipment, informatization and other scenarios that require complex logic control & Password cracking, image processing and other scenarios that are independent and require a lot of computing power &  AI voice recognition, AI face recognition and other scenarios that require deep learning \\ \hline 
\end{tabular}
\label{table31}
\end{table*}

The main differences among above three computing power are summarized in Table \ref{table31} for more clear understanding.

Apart from computing power measurement objectives above  narrowly, there are also many computing power measurement objectives broadly for researchers to explore. The broad computing power measurement include traditional hardware chips computation capability above  and computing power service ability as follows.

In addition to computing resource, computing power service is also supported by cache resource, network resource and service resource. Therefore, computing power for computing power resource pool is also measured by cache capability, network performance, microservices in upper encapsulation, latency, encoding-decoding ability, frames per second (FPS) and so on. For AI applications, the inference delay of the model is also a key measurement objective  of computing power. The lower the inference latency, the better the user experience, and the higher latency  may cause some delay insensitive  applications can not be processed successfully. Network performance is also a key objective specially for some real-time applications. Therefore, network packet delay variation (PDV) is also used to measure computing power. In the process of training deep learning model, a  key metric is the amount of data that the model can input and output per second. Therefore, throughput is a key objective for computing power. FPS refers to  the number of frames transmitted per second. The more frames processors  can handle per second, the smoother the picture becomes. With constant image resolution, the higher the GPU computing  power, the higher the FPS. 

\subsection{infrastructure computing power modeling }\label{s32}
Computing infrastructures  in computing power network include CPU  under traditional X86 universal server architecture, GPU parallel computing chips suitable for processing task in which the type of data is unified such as images processing, TPU and NPU with professional acceleration processing for neural network, ARM widely used for embedded edge devices, FPGA with programmable logic and  high computational efficiency. Moreover, some hardware chips with atomic computing function will be produced in the future. Existing computing power infrastructures (CPU, GPU, DPU, NPU, TPU, FPGA, etc.) often employ various different architectures with different computing power.  Chips produced by different manufacturers have different designs.  Moreover, chips also  have different designs for different types of computing. Therefore, it is necessary to model the heterogeneous computing power uniformly. Different types of computing power resources need to be mapped to a unified dimension. Thus, computing power resource pool formed by scattered infrastructures can be read and understood by the service layer, which can shield the heterogeneity of the underlying physical resources from the service. 

In \cite{321}, the author proposed a computing power quantity modeling as follows. Assuming there exist $n$ chips with logical computing power, $m$ chips with parallel computing power and $p$ chips with neural network computing power in a computing power resource pool. Thus, the total computing power in the computing power pool is, 
\begin{equation}\label{eq321}
C = \left\{ {\begin{array}{*{20}{c}}
{\sum\limits_{i = 1}^n {{\alpha _i}f\left( {{a_i}} \right) + {q_1}({\rm{TOPS   }})} }\\
{\sum\limits_{j = 1}^m {{\beta _j}f\left( {{b_j}} \right) + {q_2}({\rm{FLOPS}})} }\\
{\sum\limits_{k = 1}^p {{\gamma _k}f\left( {{c_k}} \right) + {q_3}({\rm{FLOPS}})} }
\end{array}} \right..
\end{equation}
In Eq. \eqref{eq321}, $C$ is the quantized total computing power. $f(x)$ is mapping function. $\alpha$, $\beta$, $\gamma$ are the mapping scaling coefficients. $q$ is the redundant computing power. For every kind of special hardware chip,  $\alpha$, $\beta$, $\gamma$ and $f(x)$ are usually given by manufacturers. Take the GTX680 an example \cite{322},  a single-core processor can perform  single-precision calculations twice in a  clock cycle. The number of the processors is 1536 and the main frequency is 1006MHz. Therefore, the peak computing power is $ P=2\times 1536 \times 1006\text{ MHz}=3.09\text{TFLOPS}$. Therefore, to calculate the total OPS or FLOPS of a computing power pool, $q$ needs to be modeled. The value of $q$ can be obtained by numerical simulation with collected historical average computing power peak value of the computing power resource pool. Assuming historical average computing power peak value as $H_c$,  thus, $q_1=H_c-{\sum\limits_{i = 1}^n {{\alpha _i}f\left( {{a_i}} \right) } }$.

Apart from using the way of numerical simulation, computing power resource pool can be also served as a black box. The historical computing power data of same type of computing power resource pools can be collected with daily tests in the research. With the development of AI and big data analysis, it is possible to measure and quantify the computing power of the computing power resource pool by deep learning algorithm. Latency and throughput, etc., provided by a computing power resource pool can be also modeled and measured by the way above.

\subsection{task requirements computing power modeling }\label{s33}
Building a customers computing power demand measurement model can map the customers task demand into the corresponding actual required computing power. Thus, computing power network system can more efficiently perceive the demand of customers more properly allocate the computing power.  The computing power demand of a computation task is related to the type of traffic such as face recognition or self-driving, the type of computation such as text or video, the size of computation data, the tolerate processing latency of computation task etc.

However, it is very difficult to compute the processing latency of computing tasks even through with exact number of operations.  What's worse, computing and storage resources of computing nodes are time-varying. Fortunately, the latency of computing tasks can be tested with different computing node condition. Therefore, the computing power demand of customers tasks can be mapped into a function as follows.

\begin{equation}\label{e431}
    \tau  = f(P,S), 
\end{equation}

where $\tau$ represents the processing latency of computing tasks, $P$ represents the performance of computing nodes including computing resources, storage resources and I/O resources, $S$ represents the properties of computing tasks. In details,
\begin{equation}\label{e432}
    P = ([{F_w},{F_0},\bar F],[{C_w},{C_0},\bar C],[{E_0},\bar E]),
\end{equation}
where ${F_w},{F_0},\bar F$ represent the whole computing capability, the percentage of computing resources that are consumed when the task arrives, the average percentage of computing resources consumed when the task is executed, respectively. ${C_w},{C_0},\bar C$ represent the whole storage resources, the percentage of storage resources that are consumed when the task arrives, the average percentage of storage resources consumed when the task is executed, respectively. ${E_0},\bar E$ represent read/write rate when the task arrives, the average rate when the task is executed, respectively. Thus,
\begin{equation}\label{e43}
    {F_0} = F(t){|_{t = {t_0}}},{C_0} = C(t){|_{t = {t_0}}},{E_0} = E(t){|_{t = {t_0}}},
\end{equation}

\begin{equation}\label{e44}
    \bar F = \frac{1}{\tau }\int_{{t_0}}^{{t_0} + \tau } {F(t)} dt,
\end{equation}

\begin{equation}\label{e45}
    \bar C = \frac{1}{\tau }\int_{{t_0}}^{{t_0} + \tau } {C(t)} dt,
\end{equation}

\begin{equation}\label{e46}
    \bar E = \frac{1}{\tau }\int_{{t_0}}^{{t_0} + \tau } {E(t)} dt,
\end{equation}

And,
\begin{equation}\label{e437}
    S = ([W,D]),
\end{equation}
where $W$ is computing load of the task, and $D$ is the size of the task. Thus,
\begin{equation}\label{e438}
    W = \int_{{t_0}}^{{t_0} + \tau } {{F_w}F(t)} dt = {F_w}\int_{{t_0}}^{{t_0} + \tau } {F(t)} dt,
\end{equation}
\begin{equation}\label{e439}
    D = \int_{{t_0}}^{{t_0} + \tau } {{C_w}C(t)} dt = {C_w}\int_{{t_0}}^{{t_0} + \tau } {C(t)} dt.
\end{equation}

With the function fitting of Eq. \eqref{e431}, the relationship between the processing latency of computing task and the computing node capability can be modeled. Thus, computing power network system can allocate proper computing power intelligently for customers tasks to improve the experience of customers. 

\subsection{computing power grading   }\label{s34}
With the development of AI technology and 5G, there are emerging various intelligent applications which are computation intensive and delay insensitive. Diversified applications have different levels of demand for computing power. Therefore, it is most necessary to classify computing power according to general standard. Computing power grading is beneficial for  computing power transaction platform designers to provide selection standard of computing power. Moreover, computing power grading can make computing power network system to schedule computing power more intelligently and faster. 

Emerging intelligent applications, such as video render, face recognition, and self-driving, etc., mainly require floating point computing power. Therefore, the floating point computing power can be used as the basis of computing power grading. Computing power  can be divided into four grades as follows for current intelligent applications. The first grading is super computing power applications like video rendering with over 1 PFLOPS level request for computing power. Take VGG16 training  on the ImageNet dataset as an example, it will require 19 PFLOPS to train iterations once. The second grading is large computing power applications like DNN training with about form  10 TFLOPS to 1 PFLOPS level request for computing power. For instance, it will require about 64 TFLOPS to recognize face images with 16 channels and 300 images concurrency. The third grading is medium computing power applications like AI inferring with about from 500 GFLOPS to 10 TFLOPS level request for computing power. For example, it will require about 4 TFLOPS to recognize face images with single channel and  300 images concurrency. The lowest grading is all computing power applications like voice recognition with less 500 GFLOPS level request for computing power. Take DeepSpeech2 algorithm for speech recognition as an example, it will require 2 GFLOPS to recognize single voice.

\section{INFORMATION AWARENESS AND ANNOUNCEMENT IN Computing Power Network}\label{s4}
Information in computing power network include user side information, computing node information, and network link information. Perceiving the information and achieving real-time information announcement are  most significant for computing power on demand, efficiently and flexibly. What is available to perceive, how to perceive, and how to leverage these perceived information are surveyed in this section. In addition,  the three ways of information announcement in the whole computing power network are also introduced.  

\subsection{aware user information  }\label{s41}
User information awareness can support to achieve better mapping between user requirements and computing resources, which is very significant to schedule computing power on demand and in time. User information awareness oriented network technology has attracted much attention in recent years, such as APN6 \cite{4d1}. In detail, user information model contains user mobility information and user resource requirements information. 

\subsubsection{Mobility Awareness}\label{s411}
Mobility awareness is very significant for resource allocation in  computing power network. User location awareness is beneficial to make  the intelligent decision  of task offloading in  computing power network, such as where and how to offload. Moreover, because of the mobility of user, if the user moves out of the coverage of the base station where the edge server is deployed, the result of offloaded task cannot be returned to user successfully \cite{4111}. Therefore, it is necessary to migrate the computing task or the computing result according to user mobility among different edge servers in  computing power network. Last but not least, with the development of AI and big data analysis technology, it is possible to predict user mobility including user location, user movement  directions and user movement velocity. The prediction of user mobility can be  realized  through the acquisition and analysis of user location information, combined with related machine learning  algorithms, such as Markov chain, Hidden Markov model, Artificial neural network,  Bayesian network and so on \cite{4112}.  Communication, computing and storage resource in  computing power network can be scheduled properly according  to the predictive user mobility, which is also helpful for load balance of  computing power network.

\subsubsection{Task Requirements Awareness}\label{s412}
User task requirements awareness includes basic task features, computing resources requirements, storage resources requirements, network resources requirements and algorithm requirements elaborated in detail as follows.

\begin{itemize} 
\item[1)] Basic task features 

Basic task features contain the size of computation data, the type of computation data,  the maximum tolerate latency of the task and task flow stickiness requirements. The rising of the B5G has brought huge computation data to network edge \cite{y62}.  Perceiving the size and the type of computation data according to the type of computation task. For example, video stream applications and video processing for object recognition are belong to high-flow applications. Perceiving the maximum tolerate latency of computation task according to urgency degree of computation task. For example, road information processing is an emergent task with ultra low tolerate latency. However, in-car entertainment applications, such as cloud games, VR movies and so on, don't  have extreme requirements in task processing latency. Perceiving task flow stickiness requirements can avoid disconnection, drop the package and traffic chaos, etc., which is very important for some task flow that needs to be served at the same node.

\item[2)] Computing performance objectives 

Computing performance objectives mainly include computing capacity requirements and the type of computation task. There are three types of computing power introduced in Section \ref{s3}, i.e., logical computing power usually with the unit of TOPS, parallel computing power usually with the unit of FLOPS and neural network computing power usually with the unit of FLOPS. Therefore, the type of computing power required by  the computation task and the processing speed need to be perceived, which can help computing power scheduling model to allocate computing resources intelligently and flexibly. 

\item[3)] Storage performance objectives 

Storage performance objectives mainly include storage capacity requirements, storage ability requirements  and storage quality requirements. Storage capacity requirements awareness of the computation task can also help computing computing power scheduling to allocate storage resources intelligently and on demand, which can improve the resources  utilization of computing power network. In fact, computing and storage resources are both  indispensable for processing computation tasks. However, different computation tasks have  various combinations of requirements for both resources. Therefore, storage resources requirements awareness is as important as computing resources requirements perception. 

\item[4)] Network performance objectives 

Network performance objectives mainly include the task requirements for network transmission delay (i.e., the transmission rate of network link), edge network gate's  network interface bandwidth, network packet delay variation (PDV), data transmission security  and bit error rate. Different grading network performance requirements will have an effect on computing power scheduling path planning. 

\item[5)] Algorithm performance objectives 

Algorithm performance objectives mainly include algorithm function requirements of the task (such as image process algorithm, text process algorithm, natural language process, etc.), time complexity of the algorithm and  space complexity of the algorithm. Various algorithm performance requirements will be helpful to choose the best computing  power node to complete the computation task with proper service. For example, it requires about 2 PFLOPS to train once COCO dataset using ResNet50 algorithm, while it requires about 6 PFLOPS using VGGNet algorithm. Therefore, it requires to allocate more computing and storage resources using VGGNet algorithm than using ResNet50 algorithm. 

\end{itemize}

\subsection{aware computing node information}\label{s42}

The computing resources are heterogenous and ubiquitous in computing power  network. The computing resources  are deployed widely at smart end like smart phones and  laptops, network edge like edge gateways and 5G base stations, cloud centers with super computing power. \cite{y63} proposed the concept of edge intelligence which accelerate the process of computing power from a cloud center to ubiquitous network edges, even end devices. Meanwhile, the status of computing resources is time-varying and every computing nodes usually is with different computation load. For example, computing nodes are idle in some places. However, computing nodes are overloaded in some bustling business districts. Therefore, it is important to perceive computing node information, which enable computing resources leveraged efficiently and avoid load unbalancing in computing power network. Computing node information that needs to be perceived mainly include the location of computing nodes, the computing resources status, the storage resources status, the bandwidth of network interface, available service, service ID and service LOCATOR,  and the load of every computing node.

Computing resources information  of computing nodes include the type of computing power, the whole computing capacity and the residual computing resources. There are many hardware chips to provide computing power, such as CPU, GPU, TPU, NPU, and FPGA, etc. For example, we usually use the number of CPU cores and CPU main frequency to measure computation ability of CPU with Million Instructions Per Second (MIPS). Moreover, computation ability of NPU for AI computation is usually measured with Tera Operations Per Second (TOPS). At the 2020 Global Artificial Intelligence and  Robot Summit hosted by the Chinese Computer Society (CCF), Huang Chang, co-founder and vice president of Horizon Robotics, proposed a new concept---Mean Accuracy-guaranteed Processing Speed (MAPS), i.e., average processing speed within a guaranteed range, to measure computation ability of NPU.  In addition to focusing on the computing power of a single kind of chip, we also need to aware the type of hardware chips. The traditional CPU is mainly used to process logical computation task, while the emerging NPU is mainly used to process AI computation tasks, such as Deep Neural Network (DNN), Convolutional Neural Network (CNN), Recurrent Neural Network (RNN), etc.  Perceiving the type of computing power and the residual computing resources of a computing power resource pooling can help to schedule computing power more efficiently. 

Storage resources information  of computing nodes mainly include storage capacity, the residual storage resources,  storage ability, and storage quality. Perceiving residual storage capacity of the edge storage servers is instructive for resource allocation in computing power network. The computation task can be offloaded to the other computing nodes or cloud center if the local computing node cannot provide sufficient storage resource. Storage ability, such as block storage, file storage and object storage, can adapt the task storage requirements. Perceiving the quality of storage resource and classifying storage resource according to quality can flexibly provide proper storage resource for computation task with the help of perceivable task requirements information. 

\subsection{aware network link information }\label{s43}
During the past several decades, with the rapid development of communication industry, it is an inevitable trend to achieve IoT or even emerging Artificial Intelligence \& Internet of Things (AIoT). In computing power network, how to fast and efficiently access computing nodes deployed ubiquitously through wireless or wired heterogenous network  and how to connect ubiquitous computing nodes to coordinate computing power are the important topics that need to be solved. 

To solve the problems above efficiently, network link information needs to be perceived dynamically and in real-time. Therefore, the real-time status  awareness of heterogenous and ubiquitous network  in computing power network system is a most key research point. Network link information in computing power network include dynamic network connection condition (i.e., whether there exists a network connection), data transmission rate, delay jitter, packet loss rate, available spectrum, whole transmit power of a 5G base station. With the network link information,  a  weighted computing power network connection graph can be mapped. The optimal transmission path for computation tasks offloading in computing power network will be efficiently and reliably selected with the weighed connection graph. 

Apart from accurate perception of user side information and computing power  network side information, research work needs to be done to schedule computing power based on predictive user information, computing node information and network link information. With the rapid development of big data analysis and deep learning technology, it is possible to predict user behavior and network traffic \cite{431}. The user requirements, computing resources status and network link condition can be predicted by information acquisition, processing and analysis with the help  of deep learning technology like DNN, LSTM and so on. The optimal resource allocation and real-time computation task scheduling will be realized in computing power network system with the support of proactive information both user side and network side. 

\subsection{information announcement in computing power network  }\label{s44}
In computing power network, different computing nodes publish their computing resources information and deployed services state information to the nearest computing routing nodes or the controllers. Then, real-time updated computing node information will be announced or reported in computing power network by computing routing nodes. There are three available architectures to announce the computing node information in the whole computing power network, i.e., centralized computing node information announcement, distributed computing node information announcement \cite{441}, and hybrid computing node information announcement. 

\begin{figure}[htbp]
\centering 
\includegraphics[width=\linewidth,scale=1.00 ]{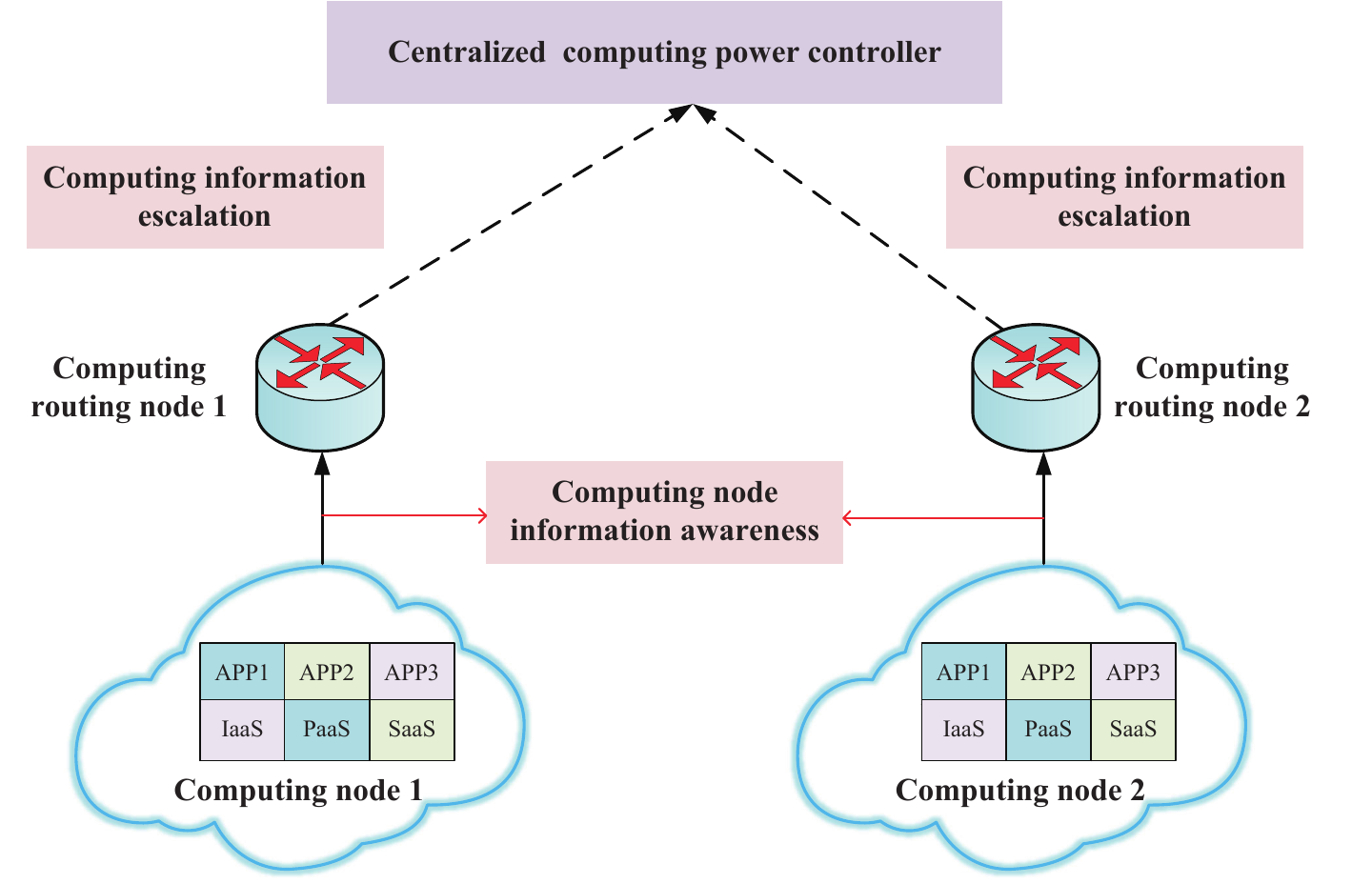}
\caption{Centralized computing node information announcement.}
\label{F441}
\end{figure}

The information of computing nodes in end-edge-cloud-network will be uniformly collected and distributed by centralized controllers such as Network Function Virtualization Orchestrator  (NFVO) as shown in Figure \ref{F441}. The  centralized controller can orchestrate computing power routing and forwarding strategies based on collected computing nodes information, and then issue to the computing power routing and forwarding nodes. The centralized controller could have a global view of computing-networking resources status of the whole computing power network. Therefore, it can achieve the routing optimization in the whole network in response to the application requests of users. Reporting the new or updated computing nodes information only involves the extension of the northbound interface, which does not need to extend the protocol. At the same time, routing and forwarding strategies are made and distributed by centralized orchestrator completely decoupling from the network forwarding devices. Therefore, the centralized computing node information announcement has the least impact on the current network devices and network protocol with the most economic cost of deployment and the shortest period of updating devices. The corresponding user requests can be accurately matched and mapped under the global computing-networking  resources view. However, computing nodes report their information to the centralized controller, which will inevitably cause the longer latency. 

\begin{figure}[htbp]
\centering 
\includegraphics[width=\linewidth,scale=1.00 ]{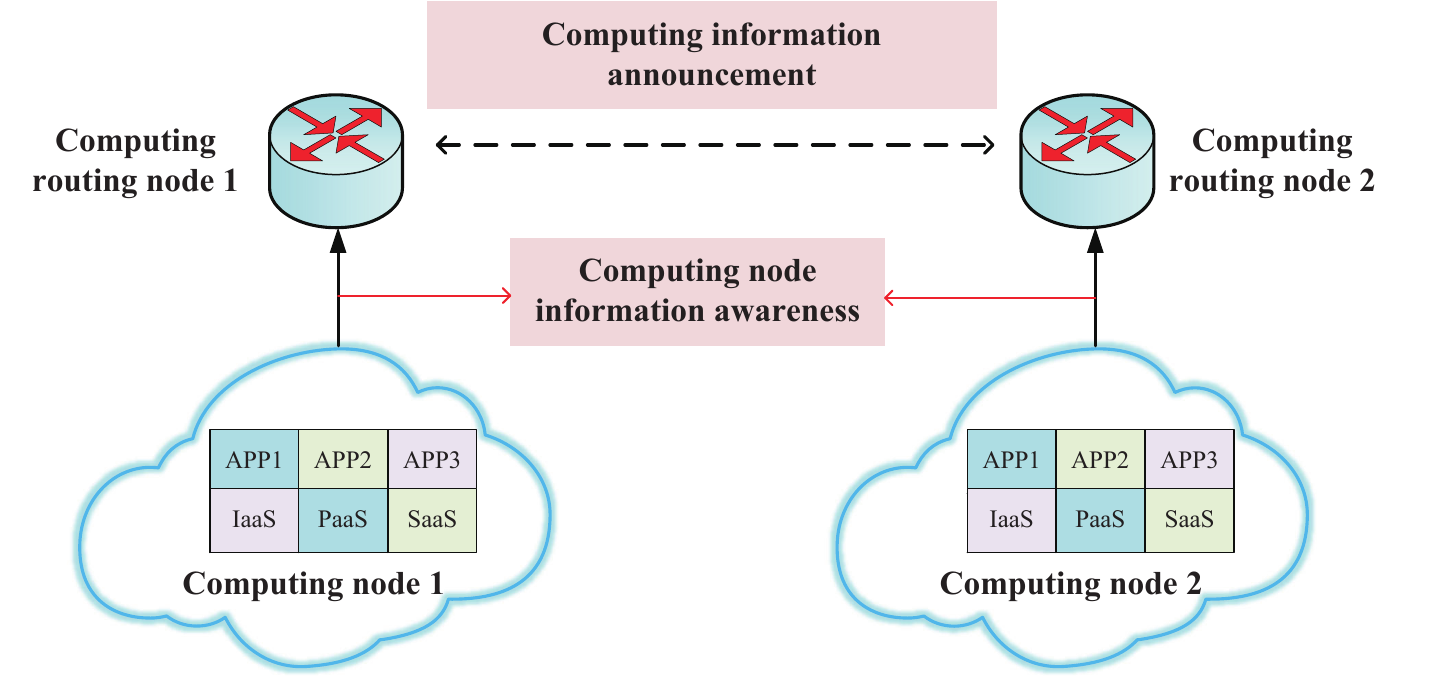}
\caption{Distributed  computing node information announcement.}
\label{F442}
\end{figure}

The computing nodes register, update or delete their information in the nearest computing routing nodes. The computing routing nodes realize computing node information announcement through Interior Gateway Protocol (IGP)\& Border Gateway Protocol (BGP) in  distributed computing node information announcement as shown in Figure \ref{F442}.  BGP can enhance the scalability of the computing power network, \cite{y61} proposed a scalable iBPG redistribution mechanism leading to optimal routing. It is unnecessary to interact with the central nodes such as the centralized controller, so the convergence speed of distributed computing node information announcement is faster with lower latency. However, applying distributed computing node information announcement need to extend the network devices in a large scale with expensive cost and a long deployment period. 

\begin{figure}[htbp]
\centering 
\includegraphics[width=\linewidth,scale=1.00 ]{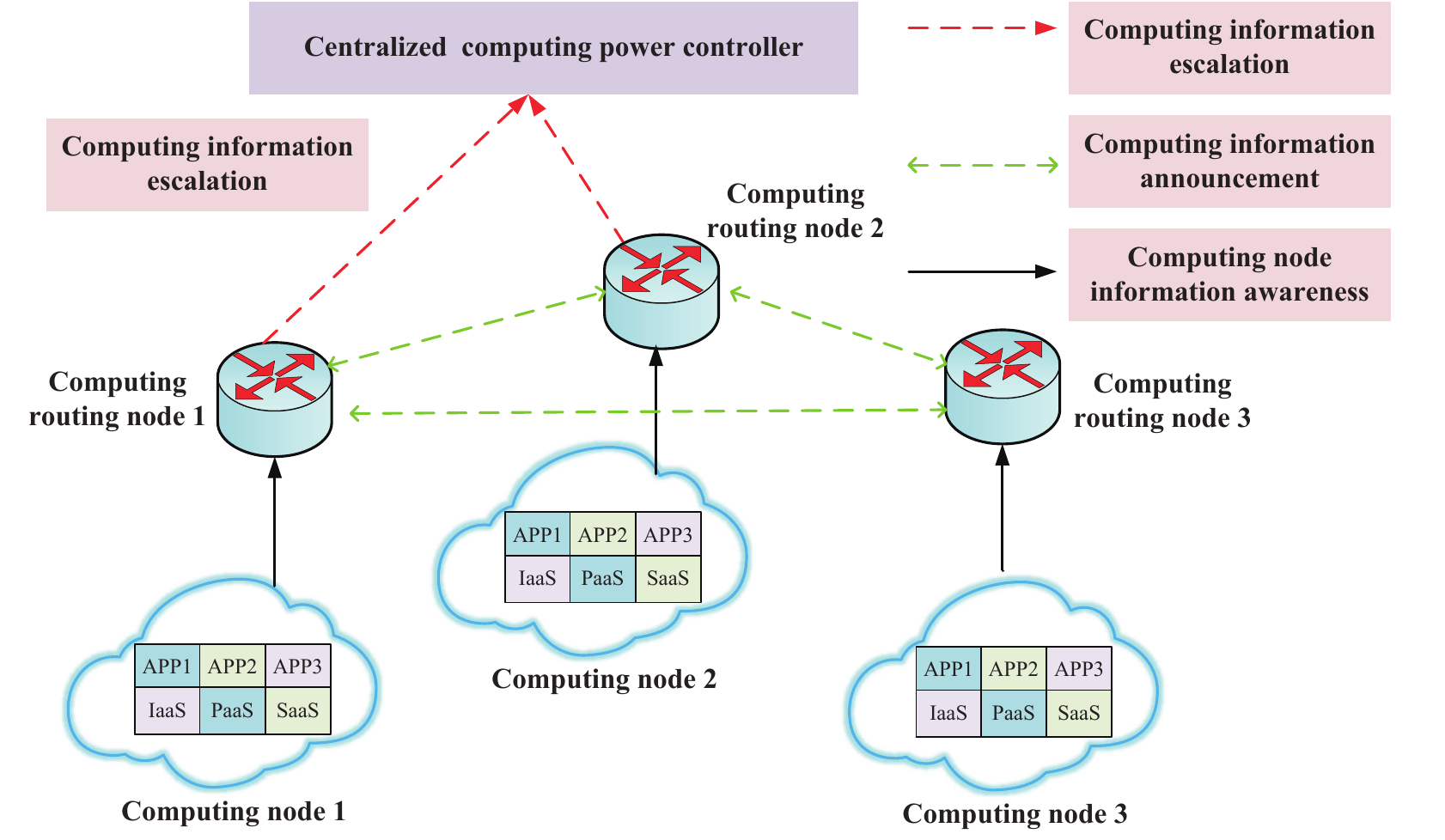}
\caption{Hybrid  computing node information announcement.}
\label{F443}
\end{figure}

Considering the advantages and shortcomings of centralized and distributed computing node information announcement, the hybrid computing node information announcement mode is proposed to balance the economic cost and response speed as shown in Figure \ref{F443}. For a large-scale computing power network, computing nodes can be divided into different groups. Computing node information announcement can be implemented by the distributed IGP\&BGP protocol within a range of the group \cite{4d2}. Meanwhile, some important computing routing nodes can report their converged computing node information to centralized controller, which will build a global resources view with shorter latency. 

In summary, a  comparison of different computing node information announcement approaches  is  summarized in Table \ref{table3}.

At present, scholars believe that there are two main ways to realize information announcement. The first mechanism is based on hierarchical extended routing. The information to be announced is divided into coarse granularity information and fine granularity information. The coarse granularity information is announced between the two edge networks based on the extended BGP protocol, while the fine granularity information is maintained by the nodes in the edge network based on the extended OSPF protocol. The second mechanism is based on new routing protocol. A new computing routing layer is designed to advertise distributed subscription or directional update based on the idea of mainstream distributed consistency protocols (such as raft \cite{4d3} / paxos \cite{4d4}). This method can decouple from the routing protocol and reduce the impact of the change of computing resources on the convergence of routing rules. In addition, we can also improve the distributed consistency protocol according to the actual network characteristics and user needs to make it more practical, such as link quality \cite{4d5,4d6}, high availability \cite{4d7}, low-delay high-throughput scenarios \cite{4d8} and so on.

\begin{table*}[htbp]
\centering
\caption{Comparison of different computing node information announcement approaches.}
\begin{tabular}{|p{5cm }|  p{3cm}| p{3cm}|p{3cm}| }\hline
\textbf{Item} & \textbf{Centralized } & \textbf{Distributed } & \textbf{Hybrid } \\\hline
Technology & NFVO & IGP\&BGP & NFVO+IGP\&BGP \\\hline
Convergence speed & Slow & Fast & Medium  \\\hline
Resource view  & Global  & Partial & Global  \\\hline
Applications-resources mapping  & Accurate & Limited & Tradeoff    \\\hline
Deployment cost  & Low    & High & Medium  \\\hline
\end{tabular}
\label{table3}
\end{table*}

\section{INTELLIGENT RESOURCE MANAGEMENT IN  Computing Power Network}\label{s5}

\subsection{computing-networking resources management   }
The research work  of computing-networking resources management mainly includes registration of computing power and computing power OAM. 

Computing resources are deployed ubiquitously in computing power network. All computing nodes need to be registered to manage computing nodes better and offload users' traffic dynamically. The registration of computing nodes includes following steps.

\noindent\textbf{Step 1:} When a computing node is ready to launch in computing power network, the computing power provider need to register the parameters information of the computing nodes in computing-networking resources management module of computing power network. 

\noindent\textbf{Step 2:} The computing-networking resources management module obtain the parameters information of the computing node consist of the type of chips and  the capacity of various resources.

\noindent\textbf{Step 3:}  The computing-networking resources management module send the computing node parameters, resource allocation and task scheduling strategies among a large number of computing nodes  to smart routers, which could store the list of computing node parameters. 

\noindent\textbf{Step 4:}  Every computing node updates their own parameters information in a serial time. The computing-networking resources management module receive the  updated parameters of the computing nodes registered in time. 

\noindent\textbf{Step 5:} The computing node will be logged out when it no longer provides computing resources for computing power network. 

The work of computing power OAM module is to monitor the computing power performance of computing resource pools by collecting the computing-networking resources information elaborated in section  \ref{s4}. If the network link status or computing resources of the currently selected computing node can not meet the service requirements, the controller need to replan the network routing path or reselect the computing nodes. With the administration of computing power OAM module, we could realize the beautiful vision of optimal computing resources scheduling, which could improve the user experience and computing-networking resource utilization. Meanwhile, computing power OAM module could detect the fault in computing power network. It could perceive the running status of computing nodes and the network link condition. The traffic could be switched to a new network link or computing nodes, once there is a problem with the network or computing node. 

\subsection{computing offloading  }
The computation task scheduling is confronted with a wide variety of challenges due to the Trilemma in computing power network. Firstly, the computing resources of edge computing nodes is usually limited. Secondly, the computing nodes located in  end devices, edge computing servers, or cloud center have usually different computing capability. Thirdly, the computing load status of the computing nodes deployed ubiquitously is dynamically variable all the time. Therefore, there exist several problems as follows that need to be solved. 

\noindent\textbf{P1:} Which computing node should be selected to execute the computation task when multiple computing nodes could complete the computation task?

\noindent\textbf{P2:} How to divide a  computation task and  offload to several different computing nodes to process coordinately when the computation task is with large computation volumes?

\noindent\textbf{P3:} When the computation task need to offload to several computing nodes inter-domain  (i.e., end-edge-cloud collaboration) to complete commonly, how to divide the computation task?

\noindent\textbf{P4:} When the computation task need to offload to several computing nodes in-domain  (i.e., edge computing nodes  collaboration) to complete commonly, how to divide the computation task and select the computing nodes?  

\subsubsection{Optimization objectives }
\begin{itemize} 
\item[1)]Minimize latency 

Latency is a key performance metric for emerging new applications which are usually computing intensive and delay insensitive. Edge computing could provide low latency but with low computing power capacity limitation. Cloud computing has high computing power capacity but with long latency because of long backhaul link.  Therefore, the combined computing of end-edge-cloud in computing power network can minimize the service latency and guarantee the computing power requirements at the same time. In \cite{5211}, the author investigated the joint service caching, computation offloading and resource allocation problem in a general multi-user multi-task scenario and formulate a mixed-integer programming problem which is non-convex and NP-hard aiming to minimize the weighted sum of all users’ computation cost and delay cost. Peng \textit{et al.} \cite{5212} optimize joint network-wide latency and energy consumption minimization problem constrained by network stability.  A distributed-two-stage offloading (DTSO) strategy  can effectively reduce latency and energy consumption and achieve a balance between them based on application preferences \cite{5214}.

\item[2)]Minimize energy consumption 

Energy efficiency is also an important metric to evaluate the performance of computing power network system. Various optimization formulations have been proposed to minimize energy consumption in the era of edge computing, which have very valuable reference for researchers to optimize resource allocation in future computing power network. Energy consumption in sensing, processing, and transmission can be  inherently integrated into the decentralized sensor selection framework \cite{5215}. 

\item[3)]Maximize resource utilization  

The unbalanced service load is a key challenge to confront in computing power network. Computing offloading is an efficient method to address this problem, which could improve computing-networking resource utilization. In \cite{5213}, this work is to maximize the completed tasks before their respective deadlines and minimize energy consumption with an end-to-end Deep Reinforcement Learning (DRL) approach to select the best edge server for offloading and allocate the optimal computational resource such that the expected long-term utility is maximized. 
\end{itemize}

In computing power network system, the computing offloading can be divided into the single-objective above and multi-objective consists of above objectives optimization according the requirements of users' applications or the system environment. 

\subsubsection{Optimization strategies}
The optimization strategies of computing offloading are a very critical research for computing power scheduling flexibly  and ubiquitous resources (such as networking resources, computing, and storage resources, etc.) integration in computing power network, which will improve the experience of computing power users and computing-networking resources utilization. 

Computing power scheduling will be performed in different locations in computing power network system, such as smart end devices, edge servers and cloud centers \cite{5221}.  Therefore, in terms of computing location, there are several computing offloading modes as follows  in computing power network. Computing offloading can be carried out from one end device to the other \cite{5222}, from end device to edge server \cite{y51}, among edge servers \cite{y53},  from end device to cloud \cite{y55}, from  edge server to  cloud \cite{y55}. Moreover, according to the characteristics of whether the computation task can be divided, computing offloading modes as follows can be carried out. At the start of the research for computing offloading, the researchers usually consider full offloading mode, i.e., the computation task is cannot divided. In \cite{y54}, the author proposed a strategy about full computing offloading. In the recent research, many researchers have started partial offloading, because some computation tasks have such  huge computing power requirements that they cannot be processed in a single computing node. Meanwhile, these computation tasks can be divided into multiple subtasks to process parallel. In  \cite{y53}, the author proposed a parallel offloading mode  of splittable tasks in heterogeneous fog networks.  Computing power network can allocate computing power dynamically and on demand by choose the most optimal computing offloading mode or the combination of several computing offloading modes. 

Some possible optimization strategies of computing offloading have been highlighted, including convex optimization, game theory and deep reinforcement learning. The computing offloading problems can be formulated into a maximization or minimization problem with some constraints such as battery, computing resources or network bandwidth, etc. Game theory is a very useful mathematical tool to study optimization strategies. Chen \textit{et al.} \cite{5223} proposed a game theory approach for efficient computation offloading decision. Zhan \textit{et al.} \cite{5222} designed a decentralized algorithm applying game theory  for computation offloading, so that users can independently choose their offloading decisions. Deep reinforcement learning is also  a promising  computing offloading optimization approach  to avoid the low efficiency of traditional optimization methods \cite{5d4}.   Wang  \textit{et al.} \cite{y52}  developed  an offloading algorithm based on DRL to optimize the problem of computing offloading and resource allocation in vehicle networks. 

\subsection{service orchestration   }
Computing power network is a new intelligent network integrating computing, storage and communication resources. The complete decoupling of the overlaying services and the underlying resources is realized based on cloud native technology. The service orchestration can be open computing-networking resources with service-oriented containers orchestration. 

At present, the research on service arrangement mainly includes inter-domain service orchestration of network and edge services, large scale orchestration of microservices and service-chaining, dynamic service orchestration and management of heterogeneous resources.

Users' requirements for services are various and changing dynamically. Therefore, dynamic activation of services deployed in containers can improve users' experience. For example, there are two kinds of available services deployed in a computing nodes but not in the same time, i.e., only one services can be activated due to limited computing resources. There is a user need two computing power services in different time.  Computing power network system can provide service for the user all the time with dynamic activation of services. When the user need service 1, computing power network service orchestration model  can activate the service 1 and close service 2. When the user need service 2, computing power service orchestration model can close service 1 and activate service 2. It is a promising approach to improve computing-networking resources utilization and realize computing resources allocation dynamically. 

\subsection{Resource management with intelligence }
Computing power network is a new network architecture that empowers AI applications. In the meanwhile, AI is also serving the resource management of computing power network. Machine learning algorithm will be deployed in computing power OAM module to monitor runtime metrics, find fault, and Detect anomalies and so on. In addition, computing power network system need to aware global computing resources information and network link information which are dynamic and time-varying. It will be very difficult to update global resources view of CPN with huge computing nodes. It will be helpful to model computing-network resources awareness based on Graph Neural Networks (GNN). What's more, deep learning algorithm such as Recurrent Neural Network (RNN), Long short-term memory (LSTM) could be used to predict the number of users' requests, which will be beneficial to orchestrate services in advance. Furthermore, there are usually huge users to compete the computing-network resources, which will lead to complex computing offloading decisions. Multi-Agent deep reinforcement learning could be applied in computing resources scheduling in CPN, which will make decisions faster with experience interacting with environment and distributed decision-making. Therefore, CPN with designed AI algorithm specially will be a self-intelligence network and no longer need manual operation and maintenance.

\section{INTELLIGENT COMPUTING POWER ROUTING AND FORWARDING IN Computing Power Network}\label{s6}
In this section, intelligent network forwarding based on segment routing IPv6 (SRv6) according to previous computing power scheduling strategies will be elaborated in detail. SRv6 \cite{6d1} is a segment routing (SR) technology based on the IPv6 \cite{6d2} forwarding plane. Combining the source routing advantages of SRH \cite{6d3}  as well as the simplicity and extensibility of IPv6. SRv6 offers a number of unique advantages such as powerful programming capabilities, ultra-simplified network structure, based on native IPv6 without changing the original IPv6 packet encapsulation structure. The main purpose of network routing in  computing power network is to solve  the problem of ``where to go " and ``what to do " \cite{6d4}. Therefore, the message header of computing power network forwarding plane need to encapsulate both IP routing and computing service routing.  

\begin{figure}[htbp]
\centering 
\includegraphics[width=\linewidth,scale=1.00 ]{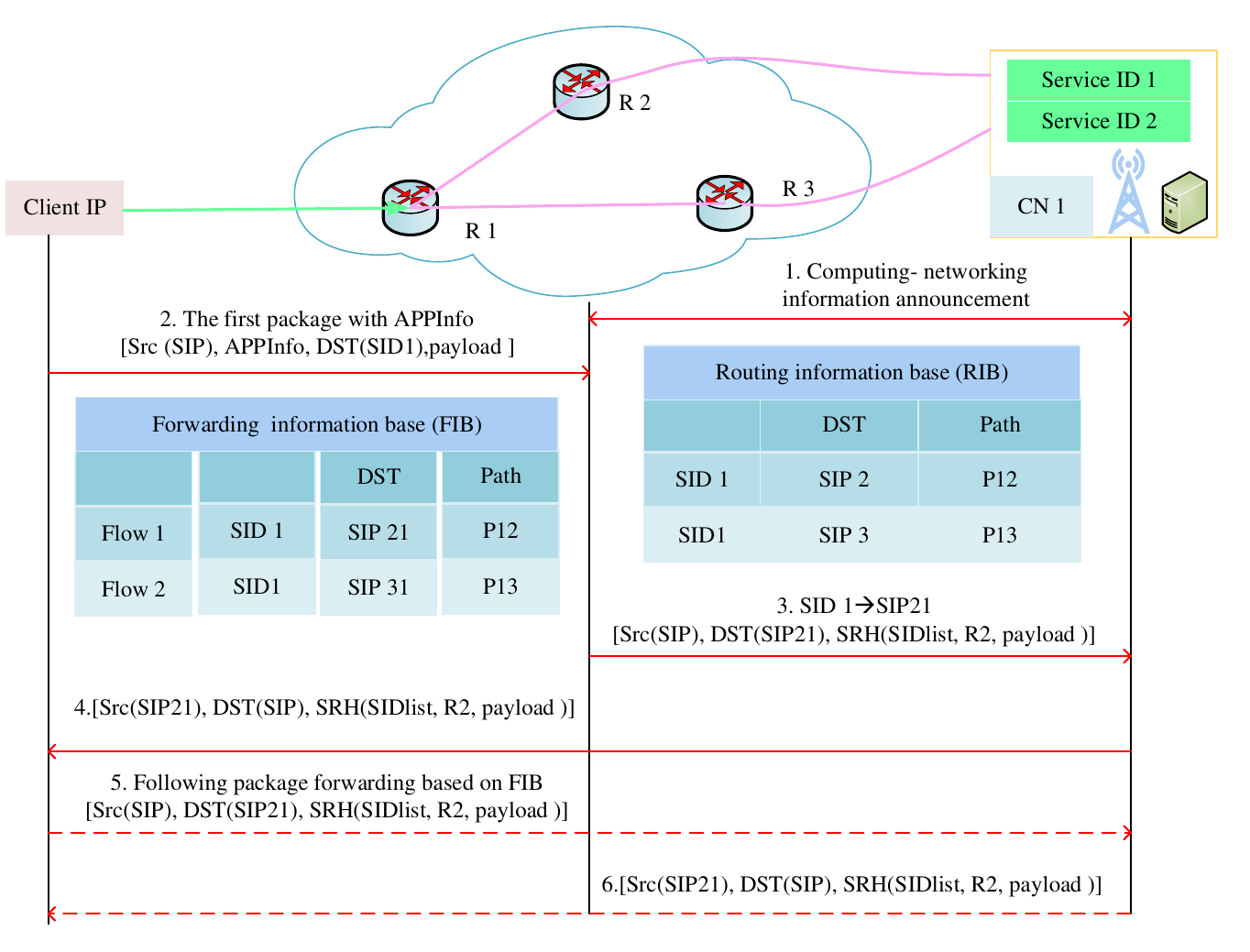}
\caption{SRv6-based computing power routing and forwarding process.}
\label{F61}
\end{figure}

The matching and mapping between applications requirements and network forwarding path will be completed at the entrance gateway \cite{6d5}. Thus, data  will be explicitly forwarded  based on SRv6. The process of  SRv6-based computing power routing and forwarding is as follows \cite{w1}, as is shown in Figure \ref{F61}.

\noindent\textbf{Step 1:} Announce computing-networking information in the whole computing power network to generate computing power routing information base. 

\noindent\textbf{Step 2:} The client sends the first package with application requirements information and destination address service ID1 to the entry node R1.

\noindent\textbf{Step 3:} The entry routing node R1 completes the mapping between service ID and service IP based the generated routing information base, assigning the exit routing node R2.

\noindent\textbf{Step 4:} The computing node processes  the client task and forwards the data according the same policy as \textbf{step 3}.

\noindent\textbf{Step 5:} Subsequent messages will be forwarded directly according to forwarding information base (FIB). 

The most key research about computing power routing and forwarding in computing power network is path selection, which will be hard to decide with traditional routing policy because there are huge connections with ubiquitous computing nodes. With the rapid development of AI technology, it will be a promising approach to selection path in CPN. First, there are many research results showing that deep learning could predict network traffic. In addition, CPN could update path selection policy dynamically  with online learning, in which every computing power routing and forwarding will provide actual data for next learning. More importantly, CPN also applies federated learning (FL) to protect data privacy. All in all, computing power network  will natively support AI capabilities to run self-intelligently.

\section{Computing Power Network PLATFORM}\label{s7}
In addition to the theoretical discussions of computing power network, the engineering implementation of computing power network for commercialization is more significant. Computing power network transaction platform and computing power network resource orchestration  platform are introduced in this section. 

\subsection{computing power network transaction platform }\label{s71}
\begin{figure}[htbp]
\centering 
\includegraphics[width=\linewidth,scale=1.00 ]{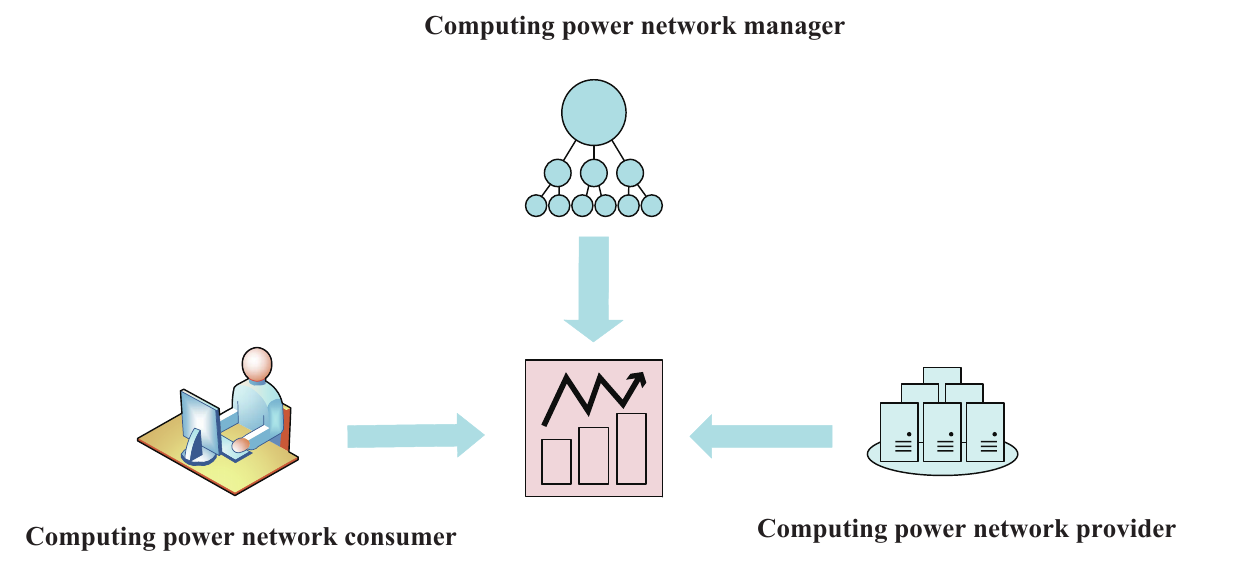}
\caption{Computing power network transaction platform.}
\label{F71}
\end{figure}

Computing power network transaction platform includes computing power network consumer, computing power network provider, and computing power network manager \cite{7d2}, as shown in Figure \ref{F71}.

Computing power network consumers request computing power resources and network resources. They request various  computing services, such as cloud games, augmented reality, face recognition, etc.,  with different performance objectives, such as latency, security, accuracy, etc.  Moreover, some computing power network consumers are mobile users specially in IoV. 

Computing power network providers can provide computing power resources, such as small-scale edge computing nodes, large-scale cloud computing nodes or even super computing centers, etc. The providers can be telecommunication operators, cloud service providers, edge cloud enterprise, or even individual devices with limited computing capability \cite{7d3}. The computing power provided by them need to be registered and authenticated first in computing power network transaction platform. 

Computing power network manager need to make computing service contract between computing power network consumers and computing power network providers. Meanwhile,  computing power billing need to be made by computing power network manager. Computing power network consumers pay bills according to the occupation statistics of computing-networking resources. Computing power network providers obtain the income according to the supplements of computing-networking resources. 

Building a trusted computing power network transaction platform based on blockchain technology \cite {10d1,10d2} is a most significant part for commercialization of computing power network. In \cite{711}, the author proposed a computing-power networking framework for ubiquitous AI by establishing 
blockchain, which is conducive to access to optimal strategies and obtain the maximum expected benefits for all parties in the network.

\subsection{computing power network resource orchestration platform }\label{s72}

The resource scheduling platform can realize flexible business scheduling and automatic service deployment according to the dependency between the network resource collection load and resources, and simplify the process of resource management while ensuring high network performance and high resource utilization. 

Some scholars have studied the resource scheduling platform in edge computing, such as the perceptible video stream analysis platform LAVEA \cite{7d1}. However, due to the shortcomings of edge computing, these studies fail to realize the combination of end edge cloud collaboration and cloud network integration. 

The basic characteristics of the resource scheduling platform based on computing power network include: general distributed computing framework, adapting to the constantly changing network environment, supporting the joint optimization of computing and resources, and better security, availability and fault recovery ability \cite{7d9}.

Obviously, these characteristics are very similar to the advantages of container cluster. Therefore, building a resource scheduling platform of computing network based on container cluster management and control applications is the future trend, and Kubernetes (K8S) \cite{7d4} is the de facto standard of container management and control. 

\begin{figure}[htbp]
\centering 
\includegraphics[width=\linewidth,scale=1.00 ]{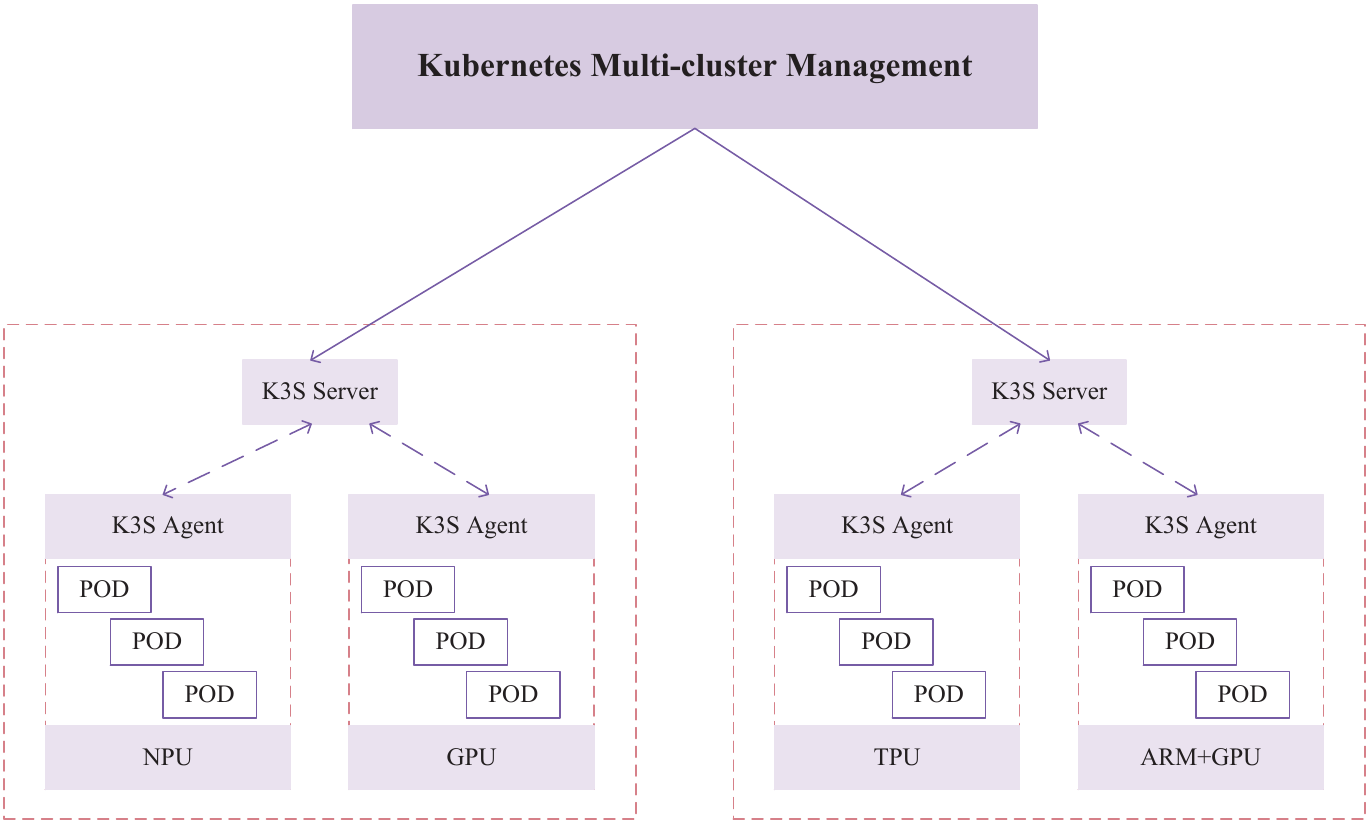}
\caption{Computing power network resource orchestration  platform.}
\label{F72}
\end{figure}

An example of computing power network resource orchestration platform applying Kubernetes is shown in Figure \ref{F72}. Nowadays, alternative Kubernetes distributions have become available, some of them specifically designed for resource-constrained environments and enables seamless extension of data and services from cloud to edge, such as KubeEdge \cite{7d5}, OpenYurt \cite{7d6}, Baetyl \cite{7d7}, and Lightweight Kubernetes (K3S) \cite{7d8}. To enable efficient service provisioning on IoT gateways, the work of \cite{722} presents the empirical evaluation of docker swarm and Kubernetes container solutions on resource constrained devices such as Raspberry Pi boards.  Therefore, given computing power scheduling in computing power network, on the one hand, Kubernetes can be applied to generally manage multiple clusters. On the other hand, K3S can be applied to realize computing resources management of limited computing nodes \cite{7d10}.

\section{COMPUTING POWER NETWORK TESTBED}\label{sA1}
To test the actual results in computing power network described above, we built a prototype testbed of CPN with key enabling technologies and several kinds of delay-sensitive and computing-intensive applications. Kubernetes is applied in our platform to flexibly deploy microservices and intelligently manage ubiquitous computing power resources\cite{liu}. We proposed a weighted directed acyclic graph model called ComNet to represent the computing resource distribution of the network, consisting of points representing computing power resources and edges of representing network link quality. Computing power awareness and announcement technology are realized for the formation of ComNet. Graph theory based computing offloading mechanism with computing modelling is designed to improve resources utilization and effectively avoid load unbalance. Several kinds of edge intelligence applications, such as face recognition, traffic detection, image augment and rendering, and distributed model training are deployed in the form of microservices to evaluate the performance of the CPN testbed.
\subsection{Computing power network overview}\label{sa1}
\begin{figure}
\centering  
\includegraphics[width=0.5\textwidth]{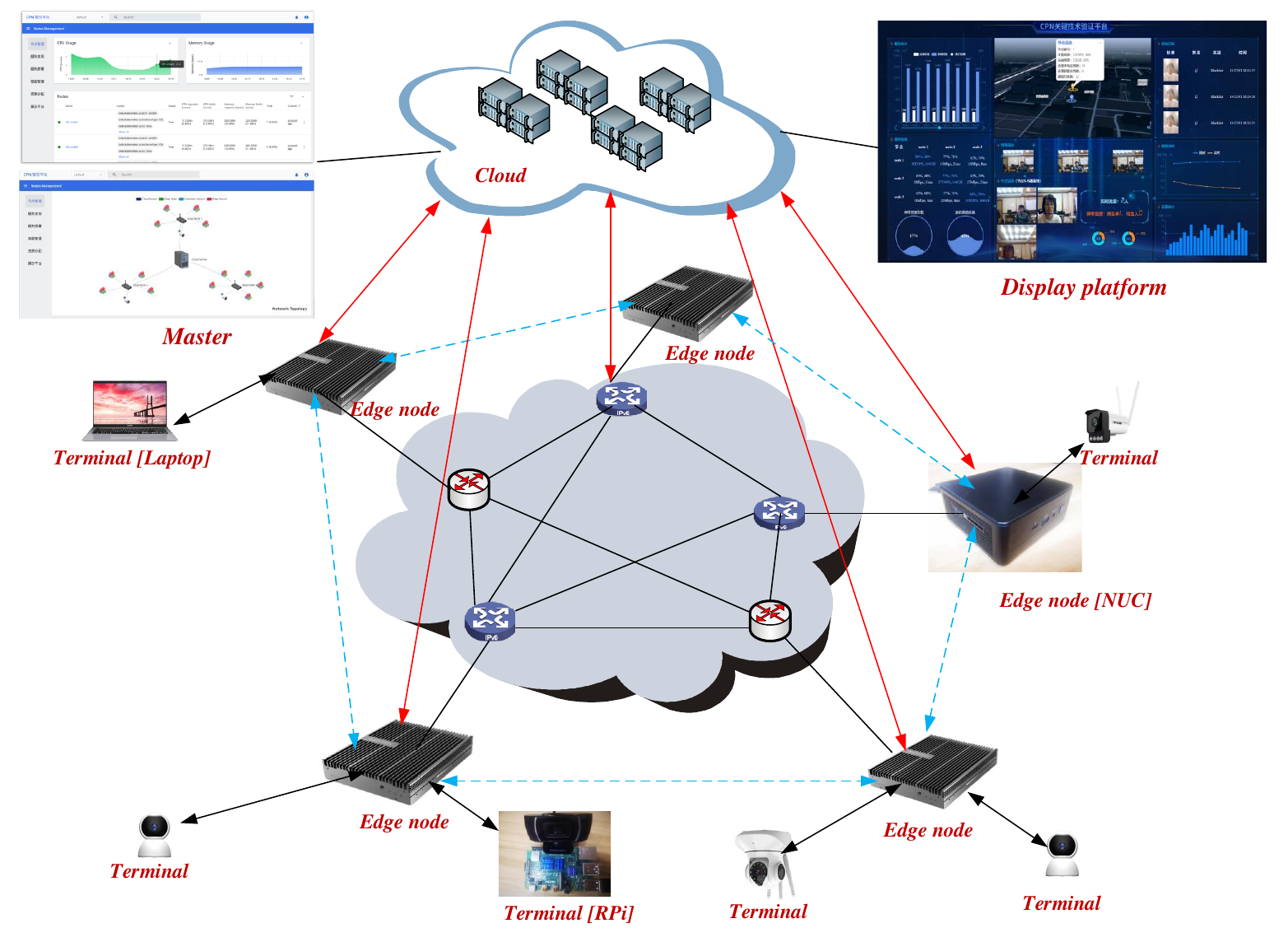}
\caption{System Architecture of Prototype Testbed.}
\vspace{-0.4cm}
\label{Fig-CPNP}
\end{figure}
\vspace{-0.1cm}
As shown in Figure \ref{Fig-CPNP}, the CPN testbed is composed of three layers: terminal, edge and cloud. The cloud is responsible for cluster management, data persistence and network performance index monitoring. The edge streamly  processes and responds to a variety of edge intelligent inference requests by terminals. The designed objectives of the testbed are: 1) Latency: the optimization goal of the testbed is to minimize the response delay of requests through computing offloading strategy. 2) Distribution: all microservices are deployed on each edge node of the cluster, where data synchronization, stable communication and load balancing strategy among diverse edge intelligence applications are all to be considered; (3) Intelligence: the platform shall realize  intelligent edge inferencing, intelligent service scheduling as well as intelligent resource management. Specifically, the main microservices of CPN testbed are as follows:

\textbf{ Computing awareness}: Based on Prometheus (official monitoring kit of Kubernetes), the microservice regularly collect the ComNet data of each edge nodes. include:
\begin{itemize}
\item Node resource: the total resources of computing and storage resource, and real-time resource utilization.
\item Link quality: link bandwidth and latency of establishing TCP connection.
\end{itemize}

\textbf{Computing announcement}: Through ZeroMQ protocol and distributed consistency algorithm, the microservice aggregates and synchronizes the other edge nodes' computing power information, and merge into the ComNet.

\textbf{Computing modeling}: We take the real-time computing power provided by each edge node as a black box. According to the ComNet and response latency data from the pre-test dataset, the neural network model is leveraged to evaluate  the computing power provided by edge nodes for several edge intelligence applications, so as to realize the service-oriented mapping from ComNet to response latency.

\textbf{Computing offloading}: Taking the minimum response delay as the optimization objective, we transform the computing offloading problem into a weighted bipartite graph matching problem, and propose an improved graph theory based algorithm (GT) to realize the multi-to-multi matching from requests to edge nodes. A mechanism for adaptive algorithm selection is also designed to achieve the optimal computing offloading performance with different network status.

\subsection{Computing power network test results}\label{sa2}
 We built a prototype testbed composed of four edge computing nodes (Intel NUC mini computer). The terminal is a Laptop computer or raspberry PI connected with several USB cameras or  IP cameras. After preprocessing the multi-channel video stream, the terminal  uploads the task data to the edge node for processing based on gRPC protocol. Various experiments are conducted to measure the platform's ability to flexibly handle a variety of intelligent reasoning services. We use the following scheduling strategies commonly used in traditional edge computing paradigms for comparison:
 
\begin{itemize}
\item Local processing (LC): Each edge node processes all arrived computing tasks locally.
\item Round-robin scheduling (RR): Each access node polls and schedules arrived tasks to other nodes in  network.
\item Greedy scheduling (GR): With computing modelling technology equipped, the arrived task is scheduled to the edge node with shortest response latency.
\end{itemize}

\begin{figure}
\centering    
\subfloat[Average responding latency.] {
\label{Fig-LAT}    
\includegraphics[width=0.5\textwidth]{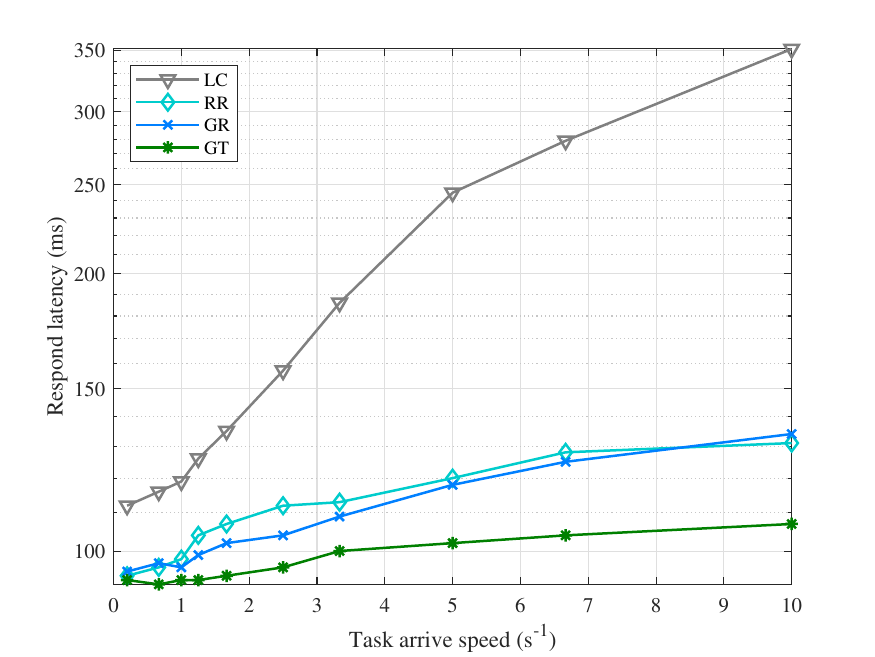}  
} 
\quad
\subfloat[Load balancing performance.] { 
\label{Fig-LB}   
\includegraphics[width=0.5\textwidth]{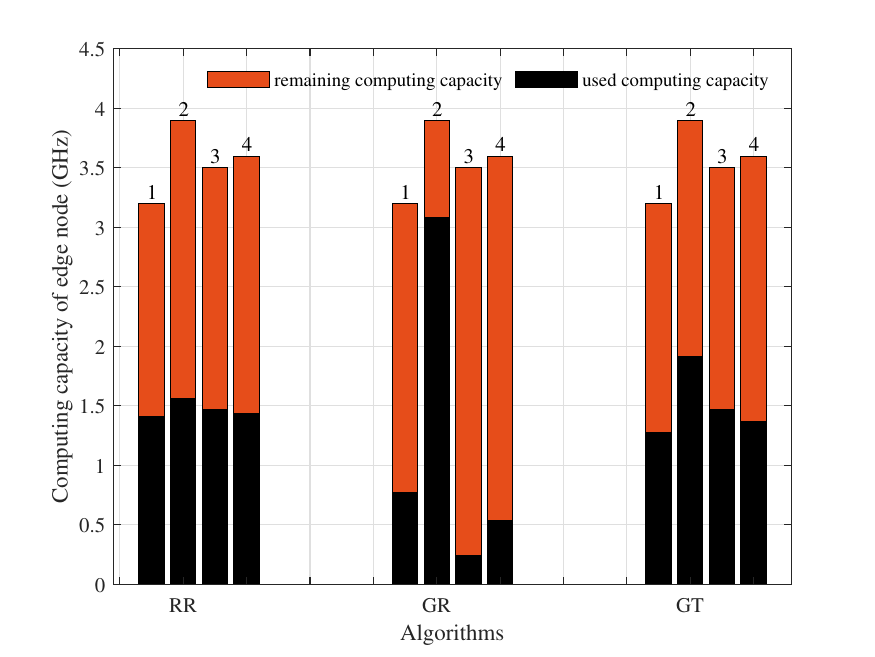}    
} 
\caption{The computing offloading performance versus the task arrival rate.}     
\label{Fig-LTC} 
\end{figure}

 We first exam the relationship between response latency and task arrival rate along with the load balance performance of each strategy under the premise of constant network topology. In this case, there is only node 2 connected to multiple devices. It can be seen in Figure \ref{Fig-LTC} that RR strategy ensures the load balance, but fails to reduce the response latency. Although GR strategy reduces the execution delay to a certain extent, it leads to the waste of resources of some edge computing nodes. Meanwhile, the proposed computing offloading strategy has excellent performance. It not only reduces the response delay by about 25\% compared with RR strategy, but also maintains a uniform node load similar to RR strategy and is much better than GR strategy.

\begin{figure}
\centering  
\includegraphics[width=0.5\textwidth]{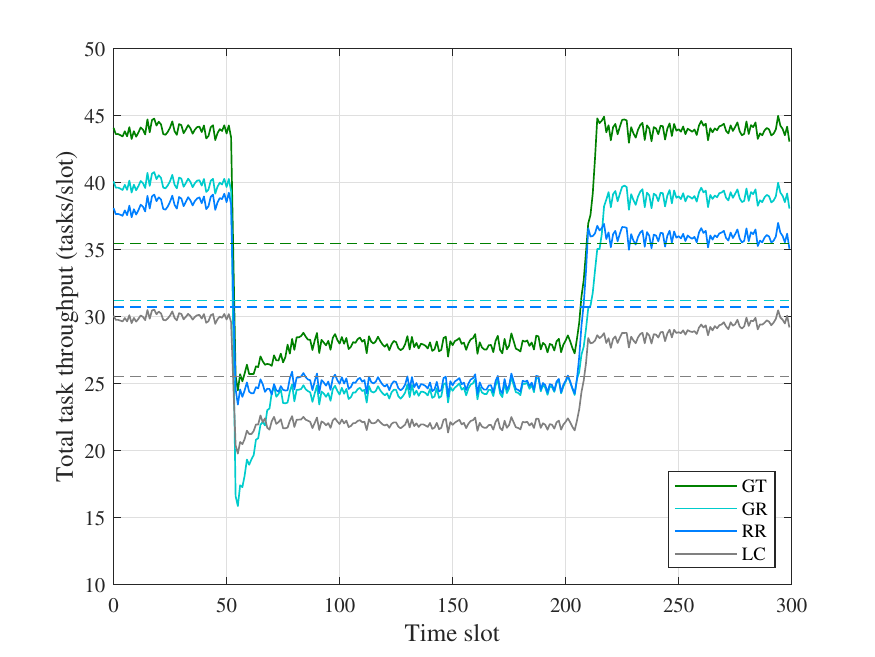}
\caption{The total task throughput versus time slots.}
\label{Fig-THP}       
\end{figure}

For evaluating the intelligence and flexibility of the CPN testbed, we test the total task throughput of all edge nodes under different computing offloading strategies as the number of edge nodes changes. In this test, node 3 is manually withdrawn (at the 50th time slot) and then rejoined to the cluster after a while (at the 200th time slot). It can be seen in Figure \ref{Fig-THP} that the task throughput recovers the original performance within less time slots after node 3 is rejoined, which indicates the computing awareness strategy we designed can smartly respond to the dynamic topology of the network and correct the decision results of the computing offloading in time. In addition, the proposed computing offloading strategy still maintains the best performance all the time. In particular, it can gradually improve the throughput by flexibly scheduling tasks after the node is disconnected. In the case with worse network quality, the performance of our testbed is still better but have not been shown due to limited space.

In our further work, more heterogeneous edge nodes will be deployed, and the distributed computing offloading strategy among multiple clusters will be realized, so as to further improve the practicability of the CPN testbed.

Notably, the CPN testbed is just the initial prototype of CPN research because we only consider to schedule distributed computing resources based on K8S without updating the network architecture. In short-term research, deploying distributed edge computing servers and managing them will be main work for computing power network. From long-term research aspect, the routing and forwarding devices in the existing network need to be updated with programmable network devices to achieve computing power scheduling in the network layer. From longer-term research aspect, the network devices need to release a portion of the computing power for computing tasks preprocessing. All in all, computing power network need to evolve gradually and update. 

\section{APPLICATIONS AND USE CASES}\label{s8}
The new computation intensive and delay sensitive applications are the main driven force for the evolution of computing power network. The requirements of emerging applications become more and more strict in latency, communication quality, computing capacity, etc. In this section, we will summarize the applications and use cases in computing power network as follows.
\subsection{Terminal Computing Power Network}\label{s80}
\begin{figure}[htbp]
\centering 
\includegraphics[width=\linewidth,scale=1.00 ]{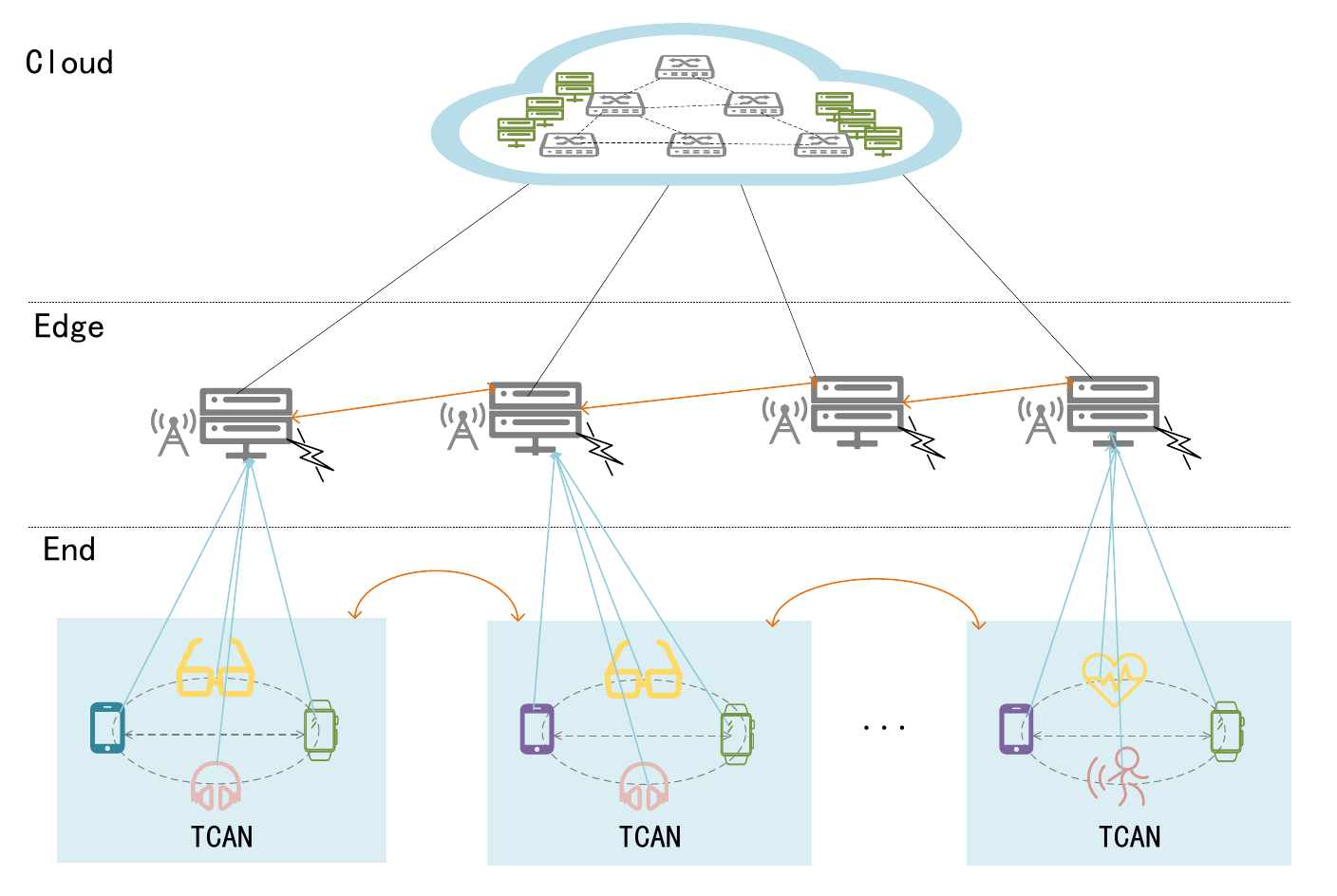}
\caption{Terminal computing power network.}
\label{F80}
\end{figure}
TCAN will include more terminals into CPN with the rapid improvement of the computing capability  of terminals as shown in Figure \ref{F80}, which will bring more spare computing power for intensive computing tasks, which is also proposed in our previous work \cite{tcpn}. The core of the TCAN is to make full use of distributed multi-level terminal computing power resources, building an efficient end-side computing power share architecture and a new network paradigm. TCAN will aiming to realize end-side computing power perception,  virtual resource collaboration, multi-granular computing power scheduling, security, privacy and trusted computing power transactions, thereby significantly enhancing the utilization ability of new computing power on the terminal side. However, there are more factors to consider because the battery capacity of the end device is limited and the end devices are more widely distributed.

In TCAN, there are several main differences from CPN. Firstly, the end devices will enter or leave the network at any time because they belong to any one rather than fixed enterprises. Secondly, short-range wireless communication between devices is usually adopted, which is usually not compatible. Finally, the end devices are mobile, which will lead  to unreliability for computing power scheduling. Although there are many challenges, it will be a promising architecture to integrate end devices into CPN in future 6G network.

\subsection{Cloud game}\label{s81}
The entertainment games are our source of happiness in the leisure time all the time. Nowadays, there are many devices which can run network games applications, such as smart phones, personal computers and laptops, etc. However, the process ability  of these smart devices is limited. Moreover, there emerges many large-scale games with the advancement of software technology and images processing ability. Therefore, the concept of cloud games is proposed in the recent years. Cloud gaming, in its simplest form, renders an interactive gaming application remotely in the cloud and streams the scenes as a video sequence back to the player over the Internet \cite{8d1}. It is necessary to transfer a cloud gaming data stream with low latency and the best possible picture quality while considering inter-dependencies  between  individual  service  provisioning chain  stages  to  ensure  user  satisfaction. \cite{8d2}

\begin{figure}[htbp]
\centering 
\includegraphics[width=\linewidth,scale=1.00 ]{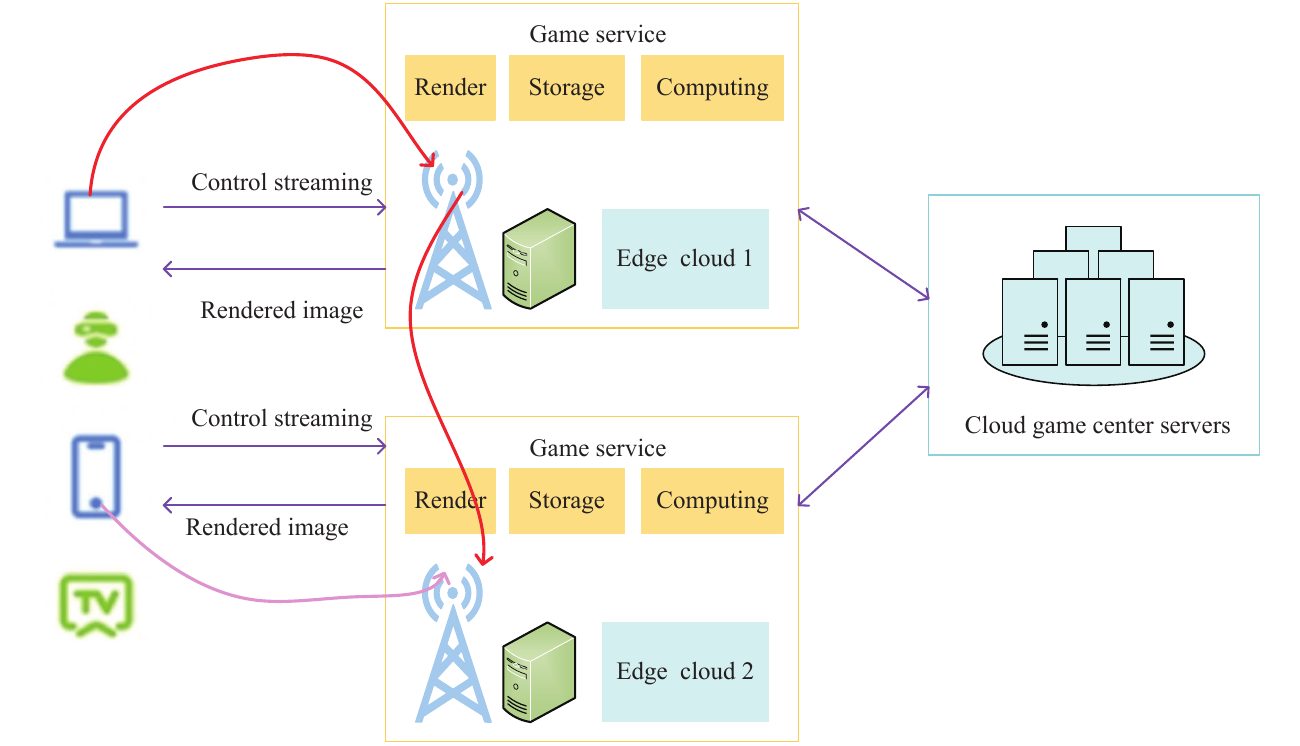}
\caption{CPN-based cloud game  system.}
\label{F81}
\end{figure}

CPN-based cloud game system is shown Figure \ref{F81}. The end devices upload  the control command to the edge cloud with image rendering, storage and computing. Then, the edge cloud will return the rendered images to end devices. Various cloud games usually need different level computing power. Therefore, a single edge cloud  completes the cloud game task computation for small-scale cloud game. Meanwhile, the edge cloud  cooperates to process the cloud game task for large-scale cloud games with CPN while cloud game task will be only processed in an isolated edge cloud with edge computing. What's more, the cloud game computation task could be scheduled to other idle computing nodes when the closest edge cloud is busy with CPN. In addition, the cloud game task could be also offloaded to nearby terminal devices with CPN without this advantage with edge computing.

\subsection{Connected vehicles }\label{s82}
Connected vehicles can communicate with different types of on-the-road devices, which goes beyond telematics, vehicle ad hoc networks, and intelligent transportation by integrating  vehicles, sensors, and mobile devices into a global network to enable various services to be delivered to vehicular and transportation systems and to people on board and around vehicles \cite{8d3,8d4}.

\begin{figure}[htbp]
\centering 
\includegraphics[width=\linewidth,scale=1.00 ]{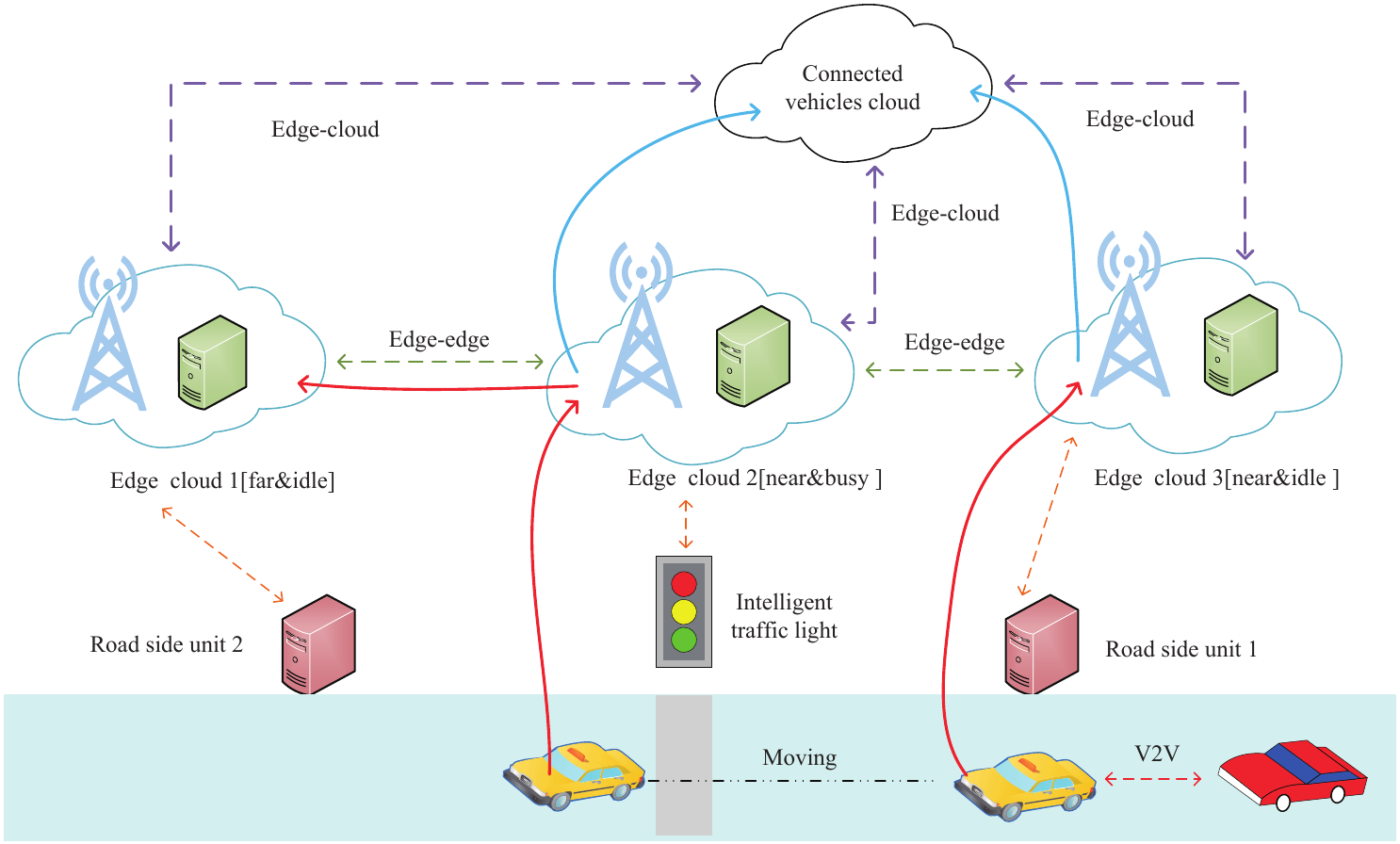}
\caption{CPN-based connected vehicle system.}
\label{F82}
\end{figure}

Computing power network plays  an important role in connected  vehicles. Assisted driving and in-car entertainment applications can be achieved  since computing power network can provide enough computing power with low latency. Moreover, traffic condition  can also be controlled orderly because computing power network can analyze real-time data from smart vehicles and road side sensors installed ubiquitously. Figure \ref{F82} shows the connected vehicles system based on computing power network. The servers deployed in computing power network will aware and analyze the data from intelligent traffic lights, smart vehicles and road side units. The  yellow car is driving from right to left passing idle  edge cloud 3, busy  cloud 2 one by one. When it is in the right near the edge cloud 3, it can offload assisted driving task (delay insensitive ) to edge cloud 3, in-car entertainment application task (delay tolerant ) to cloud. When it arrives at the middle near the edge cloud 2, it needs to offload the assisted driving task to the edge cloud 1 because edge cloud 2 is  busy and assisted driving task is delay insensitive task. Computing power network will enable computing offloading flexibly and intelligently to boost the development of connected vehicles \cite{y64}. 

Compared with edge computing, the vehicle could also offload its computation task to nearby other vehicle with CPN with shorter delay not only edge clouds. What's more, edge clouds will interact  perceived vehicles' and environmental information with each other to make a global analysis with CPN. However, there is usually only a single edge computing server to assist vehicles in their driving with edge computing. Therefore, CPN not only provide more powerful computing resources for connected vehicles but also provide more integrated decision-making compared with edge computing.

\subsection{Smart city }\label{s83}
\begin{figure}[htbp]
\centering 
\includegraphics[width=\linewidth,scale=1.00 ]{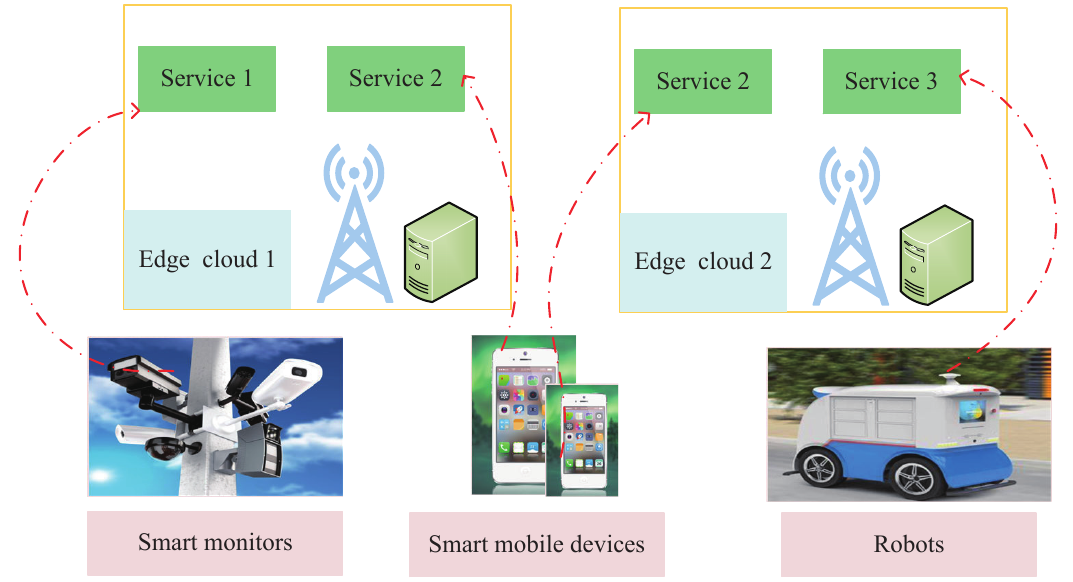}
\caption{CPN-based smart city zone.}
\label{F83}
\end{figure}

The core vision of smart city is enabling life more comfortable and secure with the advancements of many new technologies. It involves many aspects related to  our  daily life, such as smart grid, smart healthcare, smart education, smart home, smart traffic, etc \cite{8d5}. The concept of computing power network makes it nearly be in reality as shown in Figure \ref{F83}. In the city life, there are usually many computing intensive and delay insensitive applications, such as the video streaming analysis from smart monitors, delivery robots, some emerging applications on individual  smart mobile devices, etc \cite{8d6}. These smart end devices usually do not have the ability to process emerging applications. Therefore, they need to offload the computation tasks to computing power network with different service instances. The computation tasks can be mapped with special service instance in computing power network which is different from edge computing with only computing resources. Because edge computing lacks network connectivity and a mature governance platform, edge computing nodes are usually regarded as Infrastructure as a Service (IaaS). Computing power network based on Kubernetes could update deployed microservices in real time, so it is usually regarded as a fusion of Infrastructure as a Service, Platform as a Service (PaaS) and Service as a Service (SaaS). There are usually some similar services to process in smart city, such as video analysis, face recognition and so on. Therefore, CPN will bring more actual applications in smart city compared with edge computing.

\subsection{AI model training  }\label{s84}
\begin{figure}[htbp]
\centering 
\includegraphics[width=\linewidth,scale=1.00 ]{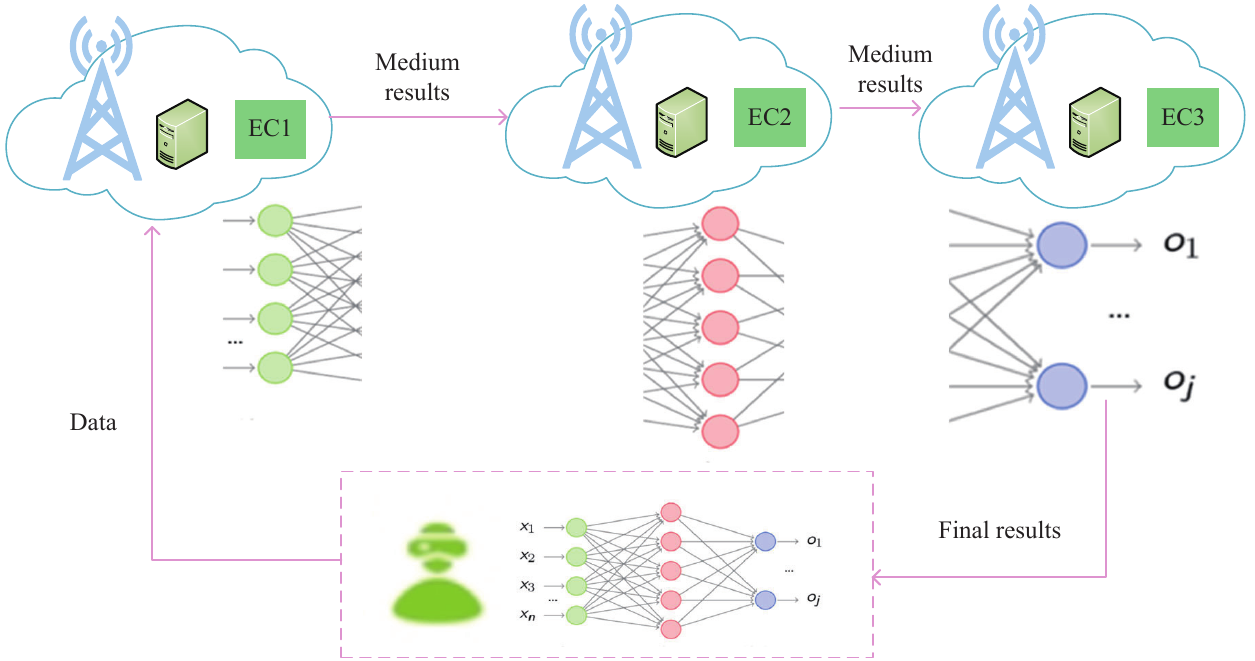}
\caption{CPN-based AI model training.}
\label{F84}
\end{figure}

In the recent years, there are emerging AI applications requesting huge computing power such as face recognition and AI-assisted medical diagnosis. These AI applications are based on trained AI model to work in real life. However, AI model training like large-scale VGGNet training need huge computing power. The computing power of smart terminal devices or single edge computing node is  usually limited. Therefore, computing nodes deployed ubiquitously need to cooperate to train the AI model as shown in Figure \ref{F84}. Most of the existing researches are based on data segmentation to realize distributed collaborative model training \cite{8d7,8d8}. In computing power network, the large-scale VGGNet can be divided into several small models while the model could be only two segments with edge computing. The trained medium results can be transmitted between the connected computing nodes. Thus, the AI model can be trained with the cooperation of  EC1, EC2 and EC3 with shorter latency. 

\section{ENABLING TECHNOLOGIES}\label{s9}
In this section, enabling technologies of  computing power network, including multi-access edge computing, software defined network, network function virtualization, artificial intelligence, sense chip, segment routing over IPv6 dataplane, microservice architecture, networking slicing are surveyed. These technologies provide intelligent management, flexibility, scalability, and security to the  computing power network.

\subsection{multi-access edge computing}\label{s91}
The concept of computing power network is a combination of multi-access edge computing and cloud computing. The advancement of multi-access edge computing makes it possible to deploy computing power infrastructures ubiquitously. Multi-access edge computing offers storage and computing resources at the edge, reducing latency for mobile end users and utilizing more efficiently the mobile backhaul and core network  \cite{911}. In \cite{912}, Apostolopoulos \textit{et al.}  proposed a novel data offloading decision-making framework, where users have the option to partially offload their data to a complex multi-access edge computing environment. He \textit{et al.} \cite{913} propose a cloud-edge  collaboration feature extraction framework, in which the feature  extraction on the low earth orbit satellite server is performed with  the assistance of the cloud server. One of the key challenges for computing power network is to efficiently collaborate multi-access edge computing and cloud computing.

\subsection{software defined network }\label{s92}
With the rapid  development of SDN technology, it is a promising way to apply SDN technology in computing power network. The core idea of SDN is to separate the control and data planes, which enable the network more intelligent, programmable, and open \cite{924}. Computing resource perceiving and computation task scheduling in computing power network can be processed centrally in SDN control planes. In \cite{921}, Zhang \textit{et al.}  constructed an SDN-assisted MEC network architecture for the vehicular network. The network state can be perceived from the global perspective, which improved the efficiency and flexibility of vehicular network. The SDN controller can sense the real-time traffic condition, which can help to achieve load-balancing \cite{922}. The  proposed SDN-MEC-VANET architecture can also  implement protocol-independent forwarding by decoupling the control plane from the data plane. SDN technology can centrally manage  the fronthaul and backhaul communication and computation resources and coordinate the cooperation between different wireless operators and technologies \cite{923}. In summary, the SDN controller can build real-time computing resource state distribution map based on computing resource state of the large distributed edge computing nodes. SDN controller would schedule the computation task to the most optimal computing nodes based on real-time computing resource distribution and network link condition when the smart terminal devices initiate computation task requirements \cite{9d1}. Applying the SDN paradigm will enable computing power network more flexible and efficient in the term of computing  resource perceiving  and computation task offloading.

\subsection{network function virtualization }\label{s93}
The core idea of network function virtualization is to virtualize a set of network functions from special hardware to general purpose computing platforms with some software technologies, which can provide the same services as traditional network  with general hardware.  The architecture of  network function virtualization management and orchestration  (NFV MANO) and MEC has been defined by European Telecommunication Standards Institute (ETSI). Computing power network can realize flexible computing power orchestration and computing power scheduling on demand with NFV. There are many research works studied about NFV orchestration.  Sarrigiannis \textit{et al.} \cite{932} proposed two novel algorithms for the joint orchestration of the MEC and cloud resources, thus enhancing the NFVO capabilities, which would be beneficial for future computing power network management and orchestration. Kiran \textit{et al.} \cite{931} formulate and solve the virtual network function (VNF) placement and resource allocation (VNFPRA) problem with a novel SDN/NFV-enabled MEC infrastructure for low-cost MEC resource allocation framework. In \cite{933}, the author proposed a  MEC decision model based on SDN and NFV technology and the solving process of the MEC decision model based on SDN and NFV.

\subsection{network slicing }\label{s94}
According to 3GPP standard, network slicing is the set of network functions and resources such as computing, storage, networking resources. Each network slicing represents end-to-end logical network implemented using dedicated virtualized resources over a shared infrastructure, which is suitable for a special application. Network slicing technology can make computing power network provide slicing service for users. The bandwidth of network slicing is flexible and dynamically adjustable. Meanwhile, network slicing can provide end-to-end various isolated services for customers based on a wide variety of network isolation technology, which can guarantee the network performance (such as latency, network bandwidth) is not affected by other network slicing. The application of network slicing to MEC systems has aroused great interests from both academia and industry. \cite{941}  proposed a novel framework for network
slicing in MEC systems, considering  the operator’s revenue  while considering traffic variations. Network slicing is a promising way to share the computing-networking resources on the same infrastructures in computing power network \cite{942,943}.

\subsection{Docker container }\label{s95}
Various computing services are placed in CPN services  layer as shown in Figure ~\ref{F21}.  Microservice  architecture in computing power network can provide service decomposition and service scheduling based on the docker container technology and container orchestration. Docker container technology can efficiently divide the computing resources of a single operating system into isolated groups \cite{951}, which can address the problem of conflicting resource usage requirements. AI services can be decomposed multiple fine-grained microservices by splitting the service vertically and the function horizontally. Therefore, various microservices are decoupled from each other, which makes it possible to use docker container technology to manage and deploy these microservices. Container orchestration automates the provisioning, deployment, networking, scaling, availability, and lifecycle management of containers. K8S is the most popular container orchestration platform \cite{952}.  Other container orchestration tools include Docker Swarm and Apache Mesos \cite{953}. Container orchestration is used to make multiple uncoupled container components cooperate with each other in computing power network. 

\subsection{segment routing IPv6 }\label{s96}
The segment routing over IPv6  data plane (SRv6) technology is a promising approach to enable services like service function chaining (SFC), Traffic Engineering (TE), and virtual private network (VPN) in backbones and datacenters.  SRv6 implements network programming through flexible IPv6 extension headers  by utilizing existing IPv6 forwarding technology. Based on SRv6 in computing power network, it is beneficial for simplifying the network  structure, realizing flexible programming, which enable network deploy new business faster. It will be no longer a vision to achieve resources scheduling on demand, reliably, and flexibly. In \cite{961}, the author showed that  application-aware Generalized Segment Routing over IPv6 (G-SRv6) network can provide fine-grained traffic steering with more  economical IPv6 source routing encapsulation, effectively  supporting 
various services. Moreover, SRv6 enabled network  could be controlled with a SDN architecture which is a key enabling technology in computing power network \cite{962}. Meanwhile, there exist a lot of research works to manage SRv6 enabled network. Performance monitoring  of SRv6 network  can be supported by the cloud-native  architecture \cite{963}. The hardware support and update are also needed for the traditional routers. In \cite{964}, Ren \textit{et al.} proposed a SRv6 incremental deployment formulation  from a traffic engineering view. With the SRv6 technology applying computing power network, computing power scheduling and forwarding will be more simplified.

\subsection{artificial intelligence }\label{s97}
With the rapid development and deep integration of computation data, computing power, and algorithms, artificial intelligence (AI) technology has been more and more mutual. The purpose of computing power network is to provide flexible computing resources for AI services, such as face recognition, smart surveillance. Meanwhile, AI algorithms, such as deep Q-learning  network (DQN)\cite{972}, deep deterministic policy gradient (DDPG)\cite{973}, and federal learning (FL)\cite{971}, are optimizing computing power scheduling for computing power network. For the future of computing power network, it will be more intelligent and flexible with the progress of AI.

\section{OPEN CHALLENGES AND FUTURE DIRECTIONS}\label{s10}
Computing power network is a revolution with the development of edge computing and edge intelligence. Therefore, there are many various research challenges and opportunities exist for future research works. In this section, some challenges are discussed, and potential research directions are proposed for future computing power network. 

\subsection{computing power modeling }\label{s101}
The basic computing power measurement has been  surveyed in Section \ref{s2}. However, there is no  general standard for metric of computing power. Computing power modeling will be a key challenge to confront as the basis of resources management and computing power scheduling. More and comprehensive factors that reflect the characteristics of computing power need to be explored. More accuracy modeling function need to be fitted maybe with the help of machine learning. 

\subsection{information synchronization }\label{s102}
Computing node information can be announced through IGP\&BGP protocol. However, the scale of computing power network is limited in the current distributed computing node information  announcement. In the future research, whether it is possible to build a new layer over IP network layer to decouple information announcement and network routing still need to explore. How to design a better protocol for the purpose of information announcement in a more large-scale computing power network with faster convergence speed is also important work. 

\subsection{mobility enhancement }\label{s103}
Services such as autonomous driving and mobile cloud games usually have very high mobility. Because of the high mobility of users, allocate the same computing and networking resources during the moving will increase the physical distance between  uses' service and computing resources, which will lead to increasing the computing latency and under-utilized computing resources. Therefore, mobility enhancement considering both horizontal and vertical mobility should be implemented, which enable computing power network allocate optimal computing resources for users. Mobility enhancement is beneficial for improving the user experience and computing resources utilization \cite{4111}. 

\subsection{communication bottleneck}\label{s10A1}
Computing power network will accelerate the implementation of the vision of cloud game, connected vehicles, smart city and AI model training, etc. However, implementing these smart applications requires real-time perception and transmission of massive amounts of data in ubiquitous environments. The dramatic increase in data volume poses a new challenge to the communication capacity of computing power networks. In order to achieve effective data transmission in computing power network, it will be a key technology research to effectively compress massive  data based on information theory, encode-decode theory and machine learning theory in computing power network. What's more, designing a load balancing algorithm to solve the communication bottleneck will be very significant, which will avoid link congestion in computing power network. Finally, In-Network computing will  be a landmark research to solve communication bottleneck in computing power network. In-Network computing can make full use of  the computing resources of network devices, such as routers, switches, to process massive data in the process of communication.

\subsection{energy consumption }\label{s104}
Computing power network enables  emerging AI applications and computing intensive tasks, which will cause huge energy consumption. Therefore, how to optimize energy consumption in computing power network to decrease the economic cost is a very key problem in the large-scale deployment and development of computing power network. There are several basic research directions for decreasing energy consumption in computing power network. Hardware chips with lower energy consumption but higher computing performance are expected to be developed, which is the basis for lower energy consumption of computing power network. The computation tasks with huge computing power requirements need to be divided into several subtasks  which need to be offloaded to other computing nodes to process. Therefore, how to efficiently divide the computation task and offload the subtasks is also valuable to research  for saving energy consumption. In \cite{1041}, the author proposed a series of priority functions to minimize energy consumption of computing offloading. \cite{1042} minimized system  total energy consumption with  user's computation latency constraint, considering both partial offloading and binary offloading.

\subsection{Deterministic ultra-low latency}\label{s105}
Deterministic networking (DetNet) technology is a new network technology for ultra-low delay and ultra-high reliability \cite{1051}, which can provide deterministic network transmission services for time sensitive users' computation tasks. The computation tasks in computing power network are almost delay insensitive, which need to realize computing offloading and  make intelligent decisions with ultra-low latency. The current network transmission mode has random uncertainty. Therefore, it is so difficult to ensure the rapid and accurate transmission of command information. It is impossible to ensure the safety of pedestrian in vehicular network.  DetNet technology can realize deterministic forwarding ability of large-scale network by applying  resources reservation, cycle mapping, path binding and aggregation scheduling approaches, etc. The deterministic ultra-low latency is expected to be realized with the combination in future computing power network combined with DetNet technology. 

\subsection{security }\label{s106}
In the computing power network, users need  frequently to use  the computing resources. Therefore, how to build trusted computing power trading between computing power providers and  computing power consumers is an important topic of computing power  network research.  Users who rent the computing power need to pay to computing power providers. With the development of blockchain \cite{1061}, the rules of computing power trading and price can be deployed in the smart  contract layer of blockchain, which is beneficial for efficient and secure computing power trading. In addition, computing resources are deployed ubiquitously in emerging smart end devices, edge gateways, and cloud data centers belonging to different enterprises. The heterogenous computing nodes provided computing power can be managed with  the distributed and trusted mechanism of blockchain. Building a blockchain-based trusted  management mechanism  of ubiquitous computing power can ensure trusted access and trusted computing power service in computing power network, which is important for the security of computing power providers and computing power consumers. The work of \cite{1062} shows that blockchain can enable a new cooperation ecosystem among multiple entities with unlimited data sharing capabilities without trusting each other.

\subsection{pricing policy  }\label{s107}
In computing power network, the networking, computing, storage, services resources are allocated flexibly and dynamically according to users' demand.  The heterogenous and ubiquitous computing resources are provided by a wide variety of enterprises or individuals. Therefore, the optimal pricing policy is most different from previous network. In \cite {y1071}, the author applied  backward induction to manage the cloud/fog resource and pricing. In \cite{y65}, the author proposed an auction-based market model to allocate computing power resource efficiently with the trading between the cloud/fog computing service provider and miners. allocation. The operation profit of computing power network is also  significantly influenced by the pricing policy when the users care about the price very much and the providers care about the profit a lot.

\section{CONCLUSION}\label{s11}
This paper surveys and summarizes the research efforts made in computing power network, which is a new network paradigm scheduling computing, storage, network resources dynamically, on demand, flexibly and collaboratively. The concept and functional architecture of computing power network are presented and we elaborate every function entity layer in detail. The related issues of computing power network are discussed respectively. For computing power modeling, the modeling objectives and basic approaches are extensively survey. For information awareness and announcement, we make a detail taxonomy on what to aware and how to announce in the whole computing power network.  We elaborate the ways  resource allocation   from resources management, computing offloading and service orchestration three aspects. Next, we analyze the computing network routing and forwarding based on SRv6 and introduce the computing power network transaction platform and resource orchestration platform. Then, we introduce computing power network testbed based on key technologies above. The novel applications and use cases are the driven force of computing power network. Therefore, we summarize some applications and use cases which computing power network can enable. The realization of computing power network is inseparable from technical support. We point out and elaborate  some key enabling technologies. This new network paradigm will face many challenges and opportunities. We list some challenges and future directions of this hot topic. 

\section*{ACKNOWLEDGEMENT}
\label{ACKNOWLEDGEMENT}
This work is supported by the National Science Foundation of China under Grant 62271062 and 62071063, and by the Zhijiang Laboratory Open Project Fund 2020LCOAB01.



\newpage
\biographies
\begin{CCJNLbiography}{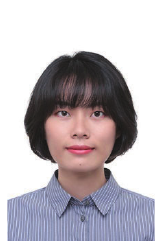}{Yukun Sun}
received the B.S. degree in communication engineering from Beijing University of Posts and Telecommunications (BUPT), Beijing, China, in 2020. She is currently pursuing the Ph.D. degree with the Key Laboratory of Universal Wireless Communication, School of Information and Communication Engineering, BUPT. Her  research interests include computing power network, computing offloading and resource allocation.
\end{CCJNLbiography}

\begin{CCJNLbiography}{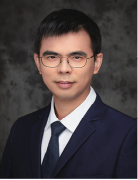}{Bo Lei}
is now a director of future network research center of China telecom research institute. Bo Lei received his Master’s degree in Telecommunication Engineering from Beijing University of Posts and Telecommunications, Beijing, P. R. China, in 2006. His currently research interests include future network architecture, new network technology, computing power network and 5G application verification. Bo Lei now leads the future network research center focusing on future network. He is the first author of two technical books and has published more than 30 papers in top journals and international conferences, and filed more than 30 patents.
\end{CCJNLbiography}

\begin{CCJNLbiography}{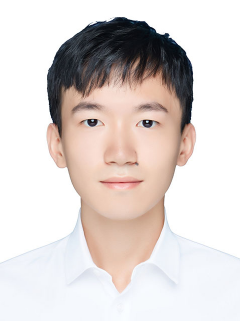}{Junlin Liu}
received the B.S. degree from Beijing University of Posts and Telecommunications, Beijing, China, in 2020, where he is currently working toward the M.E. degree at the Key Laboratory of Universal Wireless Communication. His current research interests include task offloading and resource management in mobile edge network.
\end{CCJNLbiography}

\begin{CCJNLbiography}{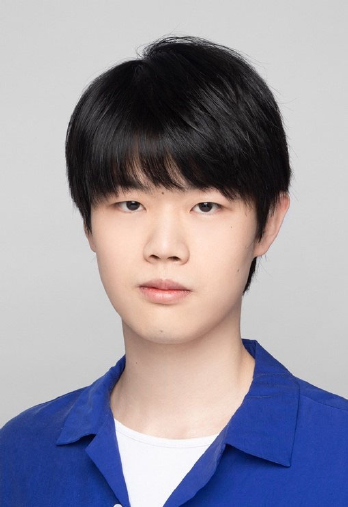}{Haonan Huang}
received the B.S. degree in Communication Engineering from Beijing University of Posts and Telecommunications (BUPT) in 2021. He is currently working towards the master degree in Information and Communication Engineering, BUPT. His current research is Edge Computing, Computing Power Network.
\end{CCJNLbiography}

\begin{CCJNLbiography}{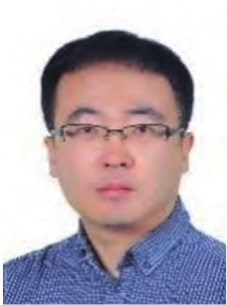}{Xing Zhang}
(M’10-SM’14) is currently  a full professor with the School of Information and Communications Engineering, Beijing University of Posts and Telecommunications, China. His research interests are mainly in 5G/6G networks, edge intelligence, and Internet of Things. He has authored or co-authored three technical books and over 200 papers in top journals and international conferences and holds over 50 patents. He received the six Best Paper Awards in international conferences. He is a Senior Member of the IEEE and a member of CCF. He has served as a General Co-Chair of the third IEEE International Conference on Smart Data (SmartData-2017), as a TPC Co-Chair/TPC Member for a number of major international conferences.
\end{CCJNLbiography}

\begin{CCJNLbiography}{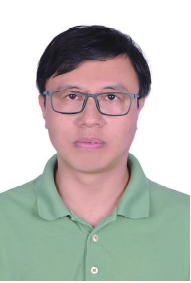}{Jing Peng}
 received M.E. degree and B.E. degree of Computer Science and Technology from Beijing University of Posts and Telecommunications, Beijing, China. He is the head of Science and Technology Innovation Department in China Telecom Beijing Branch. His areas of expertise include 5G technology, digital technology, integration of cloud and network, informatization construction and business innovation.
\end{CCJNLbiography}

\begin{CCJNLbiography}{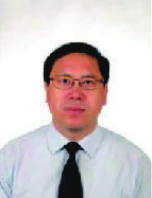}{Wenbo Wang}
received the B.S., M.S., and Ph.D. degrees from BUPT in 1986, 1989, and 1992, respectively. He is currently a Professor with the School of Information and Communications Engineering, and the Executive Vice Dean of the Graduate School, Beijing University of Posts and Telecommunications. He is currently the Assistant Director with the Key Laboratory of Universal Wireless Communication, Ministry of Education. He has authored over 200 journal and international conference papers, and six books. His current research interests include radio transmission technology, wireless network theory, and software radio technology.
\end{CCJNLbiography}
\end{document}